\documentclass[twocolumn,superscriptaddress,nofootinbib,showkeys,aps,prd,floatfix,longbibliography]{revtex4-1}

\usepackage[utf8]{inputenc}

\usepackage{footmisc,enumerate}
\usepackage{multirow}
\usepackage{subfigure}
\usepackage{amsmath}
\usepackage{amsthm}
\usepackage{physics}
\usepackage{graphicx}
\usepackage{placeins}
\usepackage{hyperref}
\hypersetup{colorlinks=true, citecolor=blue, urlcolor=blue, linkcolor=blue}
\usepackage{tabulary}
\newcommand{\orcid}[1]{\href{https://orcid.org/#1}{\,\includegraphics[width=8px]{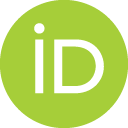}}}
\usepackage[normalem]{ulem}
\usepackage{lipsum}
\usepackage{bbold}
\usepackage[usenames,dvipsnames]{color}

\graphicspath{{./figs/}}


\begin{document}

\preprint{This article is registered under preprint number: arXiv:2302.05458;}

\title{Isospin-violating dark matter at liquid noble detectors: \\new constraints, future projections, and an exploration of target complementarity}

\author{Andrew Cheek\orcid{0000-0002-8773-831X}}
\affiliation{Astrocent, Nicolaus Copernicus Astronomical Center of the Polish Academy of Sciences, ul.Rektorska 4, 00-614 Warsaw, Poland}

\author{Darren D.\ Price\orcid{0000-0003-2750-9977}}
\author{Ellen M.\ Sandford\orcid{0000-0002-0245-8619}}
\affiliation{Department of Physics \& Astronomy, University of Manchester, Manchester M13 9PL, United Kingdom}
\date{June 8, 2023}

\begin{abstract}
There is no known reason that dark matter interactions with the Standard Model should couple to neutrons and protons in the same way. This isospin violation can have large consequences, modifying the sensitivity of existing and future direct detection experimental constraints by orders of magnitude. Previous works in the literature have focused on the zero-momentum limit which has its limitations when extending the analysis to the Non-Relativistic Effective Field Theory basis (NREFT). In this paper, we study isospin violation in a detailed manner, paying specific attention to the experimental setups of liquid noble detectors. We analyse two effective Standard Model gauge invariant models as interesting case studies as well as the more model-independent NREFT operators. This work demonstrates the high degree of complementarity between the target nuclei xenon and argon. Most notably, we show that the Standard Model gauge-invariant formulation of the standard spin-dependent interaction often generates a sizeable response from argon, a target nuclei with zero spin. This work is meant as an update and a useful reference to model builders and experimentalists.
\end{abstract}

\maketitle

\section{Introduction}
\label{sec:intro}

After decades of dedicated searches, the nature of dark matter has only been probed through its gravitational interactions. Although not necessary, many theories produce the required abundance of dark matter in the Universe through interactions with the Standard Model (SM) which are much stronger than gravity. Many such theories produce dark matter through thermal freeze-out, which has appealing features such as insensitivity to initial conditions as well as the prediction of detectable signals for experimental searches, such as colliders, telescopes, and direct dark matter detectors. So far, nature seems more complicated, and large regions of parameter space for the simplest particle models of dark matter are highly constrained~\cite{Escudero:2016gzx, Arcadi:2019lka}. 

Nevertheless, the next generation of direct dark matter detection experiments is poised to make further progress in the next decade~\cite{Aalseth:2017fik,XENON:2020kmp,DARWIN:2016hyl, LZ:2018qzl}. At the same time, the theoretical community has shown that there are numerous possibilities that are still viable. As progress is made both experimentally and theoretically, it is timely to ensure we understand any potential blind spots in our search techniques. Sometimes these blind spots are difficult to circumvent, such as $p$-wave annihilation for indirect detection searches and momentum or velocity-suppressed scatterings in direct detection. In this paper, we focus on the possibility that interactions in direct detection experiments are suppressed by a conspiracy of couplings between dark matter and quarks, such that nucleons no longer cumulatively sum, and instead cancel. This can happen when couplings between the dark matter and neutrons and protons ($c_n$ and $c_p$) are not equal and is known as isospin violation~\cite{Feng:2011vu}. 

The phenomenological motivation for considering isospin violation is clear, there are numerous theoretical models of dark matter, many of which violate isospin~\cite{Frandsen:2011cg,He:2011gc,Gao:2011ka,He:2013suk,Drozd:2015gda,Lozano:2015vlv,Chang:2017gla, Li:2019sty,Lozano:2021zbu}. 
One simple situation where the isospin-conserving case happens is when dark matter is coupled to SM quarks universally via a pure vector interaction. This is a specific case, and for a spin-1 mediated model, there are a sufficient number of free parameters to achieve any relation between $c_n$ and $c_p$, the couplings of dark matter to the neutron and proton respectively. However, for UV-consistent models that do not contain anomalies and therefore indicate some unitarity violation, the situation is often more complicated~\cite{Kahlhoefer:2015bea, DEramo:2016gos,Ismail:2016tod, Ellis:2017tkh}, leaving one with a limited number of simple models. 
Amongst them, is the $U(1)_{L_\mu-L_\tau}$ model with a fermionic dark matter candidate. Here, direct detection with nuclei proceeds from kinetic mixing between the photon and the $U(1)_{L_\mu-L_\tau}$-boson. Any vector interaction with the photon results in $c_n\neq c_p$ since their electric charges are different.

Isospin violation in direct detection has been explored previously~\cite{Tovey:2000mm,Yaguna:2016bga,Kelso:2017gib, Yaguna:2019llp}, and most commonly in the context of spin-dependent (SD) and independent (SI) interactions. Such interactions are independent of incident velocity $v$ and transfer momentum $q$ and therefore it is often a good approximation to estimate the effect of isospin violation in the $q,v \rightarrow 0$ limit. This is obviously not possible for interactions that are $v$ and $q$ dependent. In this paper, we use the non-relativistic effective field theory (NREFT) basis~\cite{Fan:2010gt,Fitzpatrick:2012ix,Anand:2013yka,Dent:2015zpa} to explore isospin-violating effects on these momentum and velocity dependent operators.

Through analysis of scintillation and ionisation atomic responses to dark matter scattering,
the liquid xenon and argon-based TPC experiments
XENON~\cite{XENON:2018voc,XENON:2020kmp,XENON:2020rca}, LZ~\cite{Mount:2017qzi,LZ:2018qzl,LZ:2021xov}, PandaX~\cite{PandaX-II:2017hlx,PandaX:2018wtu,PandaX-4T:2021bab}, DEAP-3600~\cite{DEAP:2019yzn}
and DarkSide-50/20k~\cite{Aalseth:2017fik, DarkSide:2018kuk,DarkSide-50:2022qzh, DarkSide:2022dhx}
have recently provided or will in the future provide the greatest sensitivity to dark matter for a large number of possible candidates in the GeV--TeV regime.
To study the effect of isospin violation, we carry out a realistic modeling of a representative subset of current and future xenon and argon experiments -- specifically XENON1T, DarkSide-50, DEAP-3600, LZ, and DarkSide-20k -- using publicly available data, and from this are able to reappraise the current constraints on dark matter candidates and the future prospects for detecting isospin-violating dark matter in direct detection experiments.
We evaluate the importance of certain experimental inputs such as detector efficiency on the levels of signal suppression and exemplify how the signal region can affect the interpretation of spectral shapes that arise in isospin-violating scenarios, which could lead to operator misidentification. The main results of this paper are our projections for the coming generation of argon and xenon detectors, how their limits are affected by isospin violation, and how the two targets will complement each other. We present our results by way of multiple case studies, two of which are motivated by gauge invariant dark matter models that lead to the canonical spin-independent (SI) and spin-dependent (SD) interactions. We show that, somewhat surprisingly, argon will be very competitive in the SD case. Furthermore, we present results for numerous NREFT operators in a way that accounts for the effects of $m_\chi$, such that others can use this work to estimate how other configurations of $c_n$ and $c_p$ will affect detectability.  

This paper is organised as follows, in Section~\ref{sec:NREFT_GImodels} we introduce the general formalism used to calculate predicted signals of particle dark matter in direct detection experiments as well as the two specific models we focus on. In Section~\ref{sec:exp_setup} we describe the setups we use to emulate existing and future direct detection experiments, we also describe the statistical methods we employ to generate our results, which are presented in Section~\ref{sec:results}. The results are separated into two parts, subsection~\ref{subsec:results_GI} shows our results for the spin-(in)dependent models we consider, and subsection~\ref{subsec:results_NREFT} describes how to interpret our model-independent results on the NREFT basis: a comprehensive suite of results are shown in the Appendix. Finally, we conclude in Section~\ref{sec:conclusion}.

\section{The Non-Relativistic EFT and gauge invariant models}
\label{sec:NREFT_GImodels}

In direct detection, the incident dark matter from the halo is non-relativistic and one can use the aforementioned NREFT. In this approach, one takes the non-relativistic interaction Lagrangian, 

\begin{equation}
\mathcal{L}_{\text {int }}=\sum_{N} \sum_{i} c_{i}^{N} \mathcal{O}_{i} \chi^{+} \chi^{-} N^{+} N^{-},
\label{eq:lag_nreft}
\end{equation} 
where, $\chi$ and $N$ are the dark matter and nucleon fields respectively \cite{Fitzpatrick:2012ix}. The $\mathcal{O}_i$ is a Galilean operator built from $\mathbb{1}$, the transverse initial velocity of the dark matter, $v^{\perp}$, transfer momentum, $q$, and the spin of dark matter and nucleon, $S_{\chi}$ and $S_N$ respectively. A non-exhaustive list of these operators is shown in Table~\ref{tab:operators}. The coefficients $c_i$ are model dependent and determined by the non-relativistic limit of whatever fundamental interactions take place between dark matter and quarks~\cite{DelNobile:2018dfg}. For these coefficients, we take the convention that normalises them with respect to the Higgs vacuum expectation value, $\langle v\rangle_{\rm Higgs}=264\,{\rm GeV}$, as in Refs.\,\cite{Anand:2013yka, Catena:2014epa, Bozorgnia:2018jep}. The list in Table~\ref{tab:operators} only shows operators that we explicitly consider in this work. Additional operators have been derived and studied, see Refs.\,\cite{Anand:2013yka,Dent:2015zpa,DelNobile:2018dfg,Catena:2019hzw}, but we have decided not to include them in this work. Table \ref{tab:operators} covers all operators that arise for spin-0 and spin-1/2 dark matter models with scalar, fermion, or vector mediators. For the operators above $\mathcal{O}_{11}$, known scenarios include a spin-1 dark matter candidate which is mediated by a vector or charged fermion. Often these operators are present along with those shown in table \ref{tab:operators} and are frequently subdominant. Nonetheless, they can be important for specific cases \cite{Catena:2019hzw} and a study of operators beyond $\mathcal{O}_{11}$ will be the subject of future work.

\begin{table}[th]
    \centering
    \begin{tabular*}{\columnwidth}{@{\extracolsep{\fill}}ll@{}}
    \toprule
    $\mathcal{O}_1 = \mathbb{1}_{\chi}\mathbb{1}_N$   
        & $\mathcal{O}_7 = \vec{S}_N\cdot  v^{\perp}\mathbb{1}_\chi$   \\
    $\mathcal{O}_3 = i\vec{S}_N\cdot\left(\frac{\vec{q}}{m_N}\times v^{\perp}\right)\mathbb{1}_\chi$ 
        & $\mathcal{O}_8 = \vec{S}_\chi\cdot  v^{\perp}\mathbb{1}_N$    \\
    $\mathcal{O}_4 = \vec{S}_\chi\cdot \vec{S}_N$ 
        & $\mathcal{O}_9 = i\vec{S}_\chi\cdot\left(\vec{S}_N\times\frac{\vec{q}}{m_N}\right)$  \\
    $\mathcal{O}_5 = i\vec{S}_\chi\cdot\left(\frac{\vec{q}}{m_N}\times v^{\perp}\right)\mathbb{1}_N$ 
        & $\mathcal{O}_{10} = i\vec{S}_N\cdot\frac{\vec{q}}{m_N}\mathbb{1}_\chi$  \\
    $\mathcal{O}_6 = \left(\vec{S}_\chi\cdot\frac{\vec{q}}{m_N}\right) \left(\vec{S}_N\cdot\frac{\vec{q}}{m_N}\right)$        
        &  $\mathcal{O}_{11} = i\vec{S}_\chi\cdot\frac{\vec{q}}{m_N}\mathbb{1}_N$ \\
    \hline
    \hline
    \\
    \end{tabular*}
    \caption{List of the NREFT operators constructed to obey Galilean invariance, introduced by~\cite{Fitzpatrick:2012ix}. More operators have been introduced recently~\cite{Dent:2015zpa}, but we list only operators we consider in this work for brevity.}
\label{tab:operators}
\end{table}

The differential recoil rate, w.r.t. nuclear recoil energy $E_R$ is given by,
\begin{equation}
\frac{\dd R}{\dd E_R}=\frac{\rho_{0}}{m_{T} m_{\chi}} \int_{v_{\min }} d^{3} v\, v \,f(\vec{v}) \frac{d \sigma_{\chi T}}{d E_{R}},
\label{eq:difrate}
\end{equation}
where $\rho_0$ is the local dark matter density, $m_T$ is the target nuclei mass, and $v$ is the velocity of dark matter, which is distributed according to the halo distribution $f(v)$. The integral limit $v_{\rm{min}}$ refers to the minimum velocity required to induce a recoil of energy $E_R$. The differential cross-section, $d \sigma_{\chi T}/d E_{R}$ is related to the coefficients in Eq.\,\eqref{eq:lag_nreft} by

\begin{equation}
\frac{d \sigma_{\chi T}}{d E_{R}}=\frac{m_{T}}{2 \pi} \frac{1}{v^{2}} \sum_{i j} \sum_{N, N^{\prime}=n, p} c_{i}^{N} c_{j}^{N^{\prime}} \mathcal{F}_{i, j}^{N, N^{\prime}}\left(v^{2}, q^{2}\right),
\label{eq:diff_crosssec}
\end{equation}
where $\mathcal{F}_{i, j}^{N, N^{\prime}}$ are the nuclear form factors. We follow the convention of Ref.~\cite{Fitzpatrick:2012ix}, where the operator interference is explicitly included. These are decomposed into specific nuclear response functions which have been parameterized and made publicly available~\cite{Fitzpatrick:2012ix,Anand:2013yka,Hoferichter:2018acd,Hoferichter:2019uwa}. 

Although the NREFT approach allows for a powerful model-independent analysis of direct detection data (see Refs.\,\cite{LUX:2020oan,DEAP:2020iwi,SuperCDMS:2022crd, XENON:2022avm} for recent experimental analyses), specific models only require a small number of them to be treated. Additionally, due to the hierarchy in scales for certain operator responses, and the ubiquity of some operators ($\mathcal{O}_1$, for example), not all operators are of equal interest theoretically. For this reason, we focus on two effective particle models that facilitate isospin violation in a realistic and general way. Namely, we consider a fermionic dark matter particle with either vector-like couplings ($\bar{\chi}\gamma_{\mu}\chi$) or axial vector-like couplings ($\bar{\chi}\gamma_{\mu}\gamma_5\chi$). Of course, both can be present simultaneously, but we consider them separately because the former leads to the dominant $\mathcal{O}_1$ response and if dark matter is Majorana, the later is the only non-vanishing vector current. 

We want to ensure that SM gauge invariance is always manifest, in order to do that we follow the parametrisation of Refs.\,\cite{Alanne:2022eem, Bishara:2018vix}, where we start with the relativistic effective operators

\begin{eqnarray}
     Q_{2,i}^{(6)}&=  (\bar{\chi}\gamma_\mu\chi)(\bar{Q}_L^i\gamma^{\mu}Q_L^i)\nonumber\\
     Q_{3,i}^{(6)}&=  (\bar{\chi}\gamma_\mu\chi)(\bar{u}_R^i\gamma^{\mu}u_R^i)\nonumber\\
     Q_{4,i}^{(6)}&=  (\bar{\chi}\gamma_\mu\chi)(\bar{d}_R^i\gamma^{\mu}d_R^i), 
     \label{eq:effop_BEWSB}
\end{eqnarray}

\noindent where $Q_L$, $u_R$ and $d_R$ are the quark left-handed doublet, up-type and down-type right-handed singlets respectively. The index $i$ denotes the quark generation. The operator numbering we use follows ref.~\cite{Bishara:2018vix}, which contains the dimension in the superscript (for this paper always 6), the generation $i$, and an arbitrary number that labels the operators by their form. After electroweak (EW) symmetry breaking, the three operators above match onto 

\begin{equation}
    \mathcal{Q}_{1,q}^{(6)}=(\bar{\chi}\gamma_\mu\chi)(\bar{q}\gamma^\mu q)\,\,{\rm and}\,\, \mathcal{Q}_{3,q}^{(6)}=(\bar{\chi}\gamma_\mu\chi)(\bar{q}\gamma^\mu\gamma_5 q)\label{eq:effop_AEWSB}
\end{equation}

\noindent to ensure that the Lagrangian after EW symmetry breaking remains gauge invariant under the SM, one has to ensure that the couplings in the chiral basis of Eq~(\ref{eq:effop_BEWSB}) gives the same results as that of the Dirac basis in Eq~(\ref{eq:effop_AEWSB}). The matching conditions are~\cite{DEramo:2014nmf,Bell:2015sza,DEramo:2016gos}

\begin{eqnarray}
     \mathcal{C}_{1,u}^{(6)}&=\frac{\left(C_{2,1}^{(6)}+C_{3,1}^{(6)}\right)}{2}\,\,{\rm and}\,\,\mathcal{C}_{3,u}^{(6)}=\frac{\left(C_{3,1}^{(6)}-C_{2,1}^{(6)}\right)}{2}\nonumber\\
     \mathcal{C}_{1,d}^{(6)}&=\frac{\left(C_{2,1}^{(6)}+C_{4,1}^{(6)}\right)}{2}\,\,{\rm and}\,\,\mathcal{C}_{3,d}^{(6)}=\frac{\left(C_{4,1}^{(6)}-C_{2,1}^{(6)}\right)}{2},\nonumber\\
\end{eqnarray}
where the $C$'s and $\mathcal{C}$'s represent coefficients to the effective operators before EW symmetry breaking, Eq.~(\ref{eq:effop_BEWSB}) and after, Eq.~(\ref{eq:effop_AEWSB}), respectively. With this matching, the four coefficients that are relevant to direct detection contain some redundancy. We eliminate this redundancy by following the parametrisation of Ref.\,\cite{Alanne:2022eem}, which focused on the dark axial-vector interaction. Here we do it for dark vector-like couplings as well 

\begin{eqnarray}
    C_{2,1}^{(6)}\rightarrow \frac{g^{\prime 2}}{\Lambda^2} &\cos{\theta}&,\,\,\,\, C_{3,1}^{(6)}\rightarrow \frac{g^{\prime 2}}{\Lambda^2} \sin{\theta}\cos{\phi},\nonumber\\
    &C_{4,1}^{(6)}&\rightarrow \frac{g^{\prime 2}}{\Lambda^2} \sin{\theta}\sin{\phi}.
    \label{eq:SIparam}
\end{eqnarray}

The $g^\prime$ and $\Lambda$ are introduced to make the connection with the high-scale model that produces the effective operators of Eq.~(\ref{eq:effop_BEWSB}), $g^\prime$ is the gauge coupling and $\Lambda$ is the energy scale of the new model. Since all processes in this work are assumed to be well below $\Lambda$, we treat $g^\prime/\Lambda$ as a combined parameter that sets the overall strength of interactions. The angles $\theta$ and $\phi$ generically parameterize the relative strengths of the three interactions. By adopting this parametrisation we are explicitly keeping interactions SM gauge-invariant and maintaining some connection with the full extension of the SM that explains DM.
When one performs the mapping of these coefficients to the NREFT coefficients in Eq.\,\eqref{eq:lag_nreft}, they obtain contributions to the NREFT operators $\mathcal{O}_{1}$, $\mathcal{O}_{7}$, $\mathcal{O}_{9}$. Typically because the coherent nuclear response to $\mathcal{O}_{1}$ is so large, one ignores the other contributions. Since this paper is concerned with cancellation effects, it is conceivable that there are areas of parameter space where $\mathcal{O}_{7}$ and $\mathcal{O}_{9}$ are important, perhaps limiting the isospin suppression. 

\begin{figure*}[htbp]
    \includegraphics[width=0.49\textwidth]{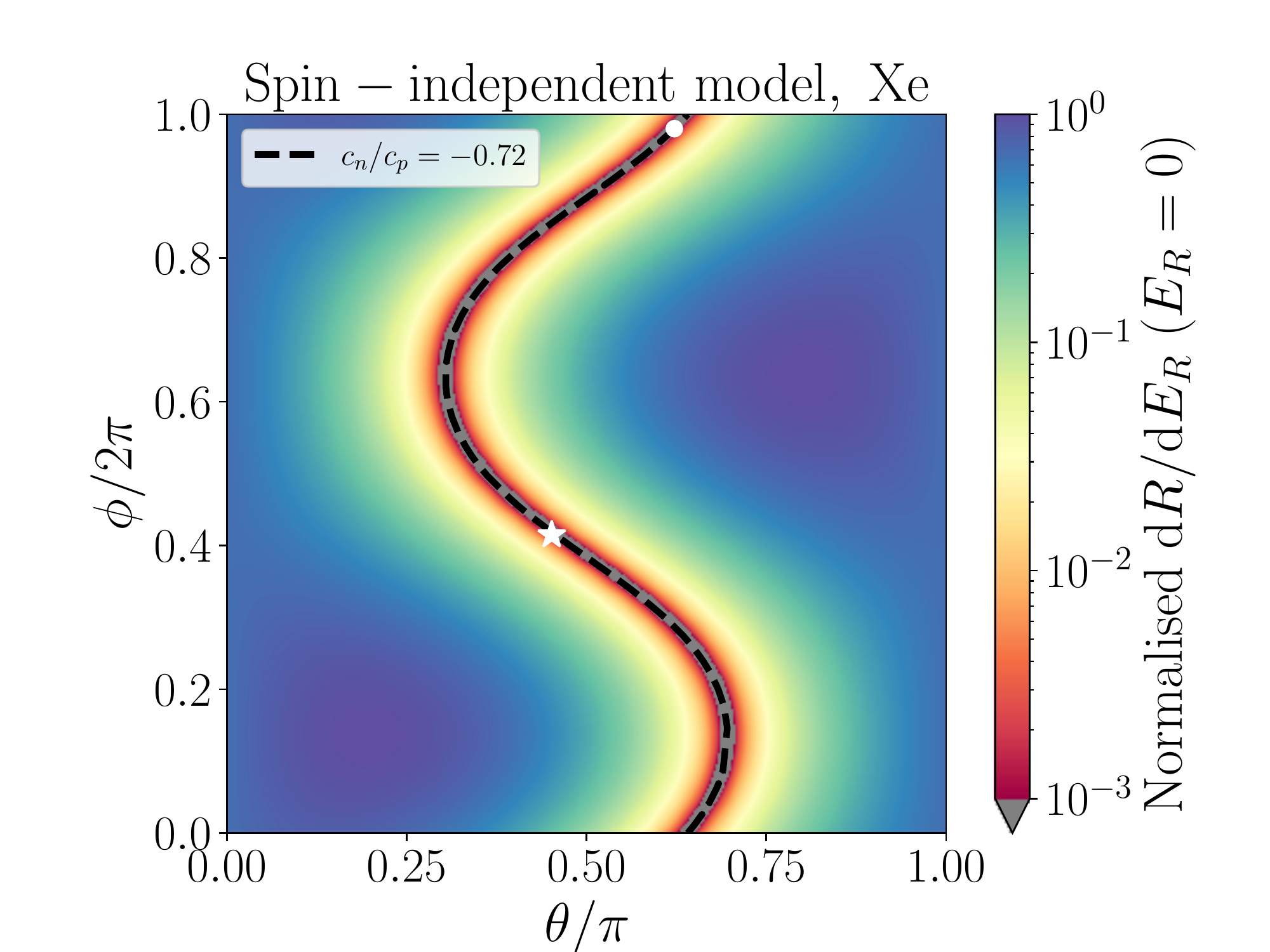}
    \includegraphics[width=0.49\textwidth]{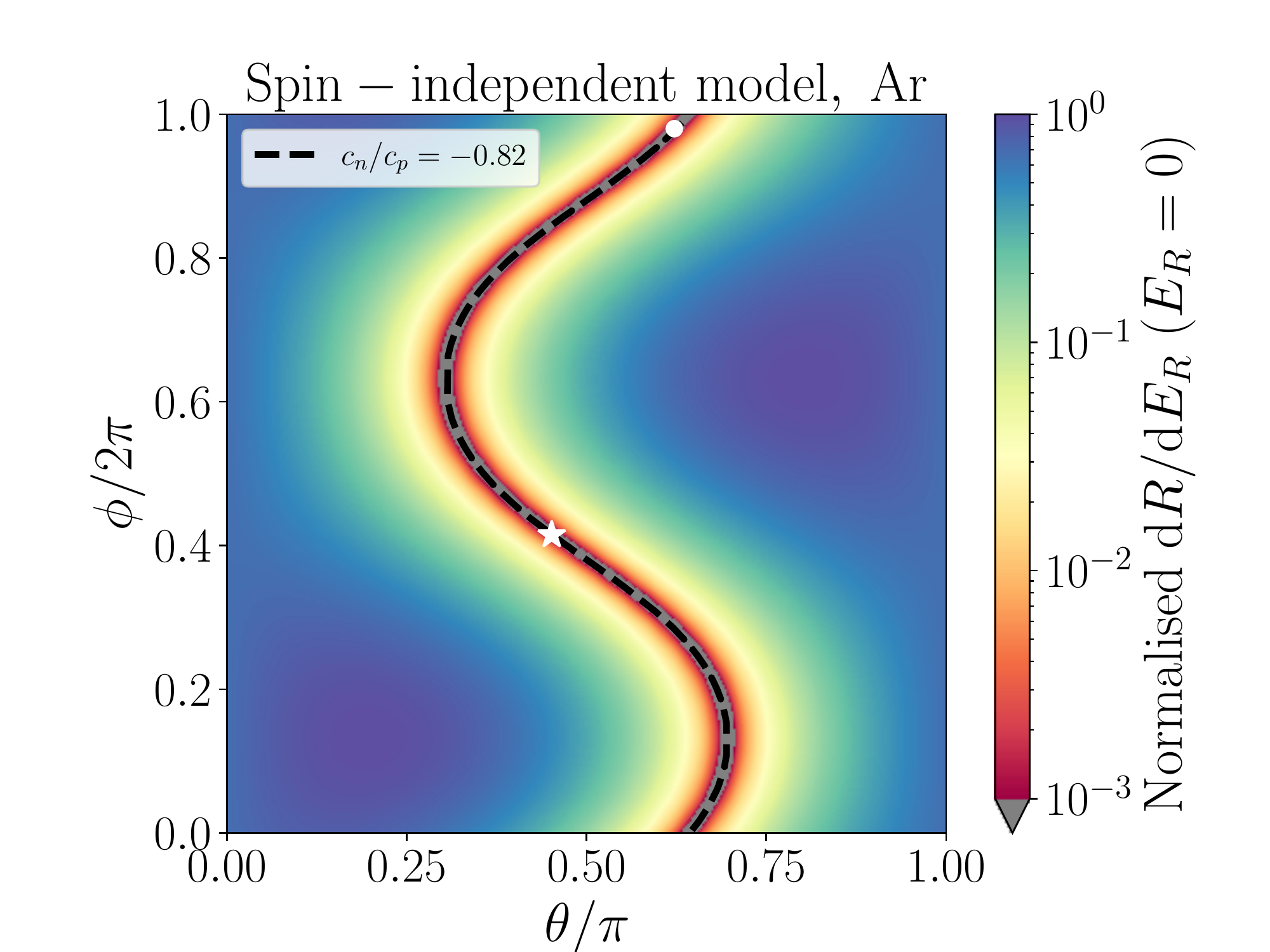}
    \includegraphics[width=0.49\textwidth]{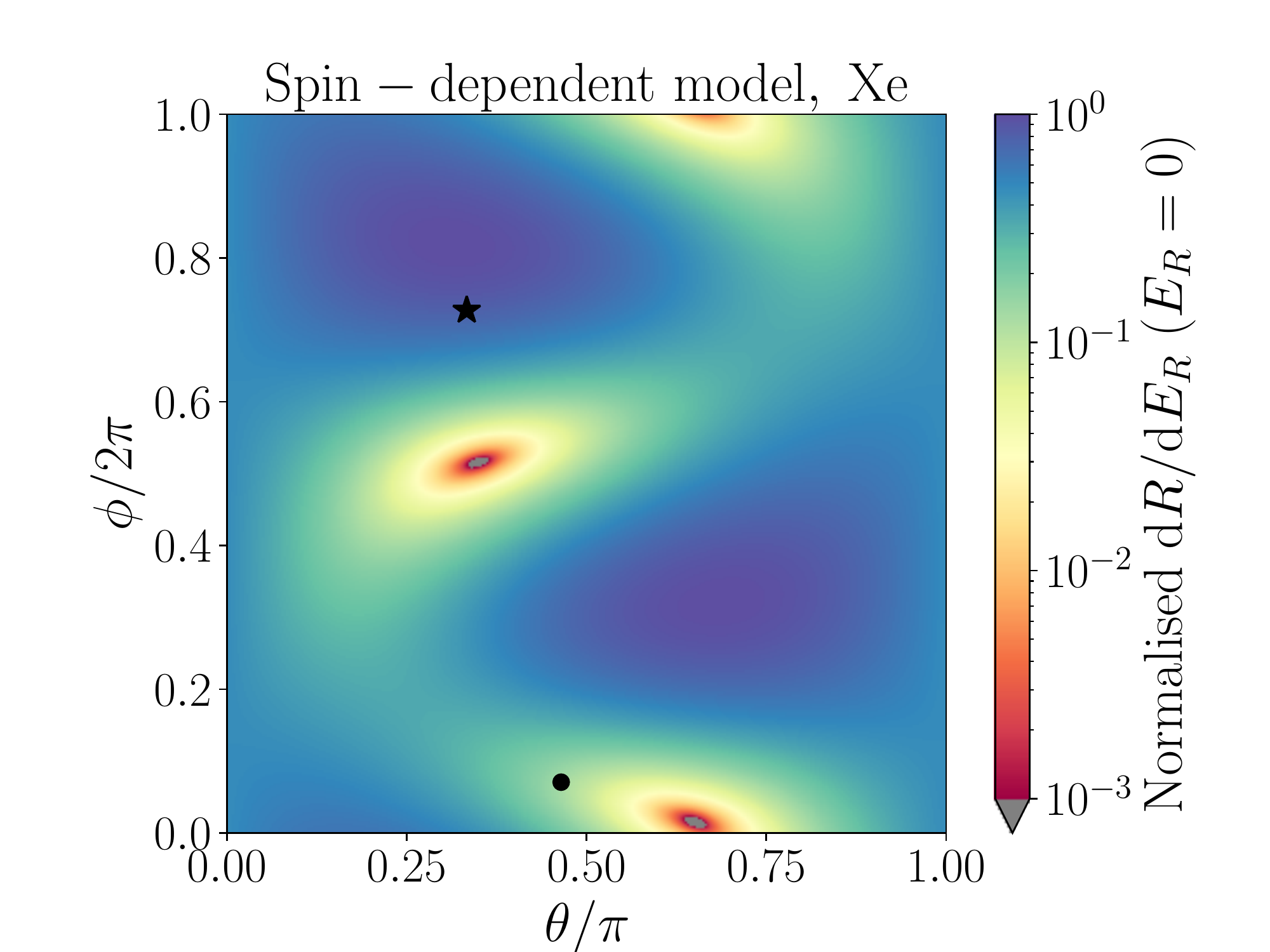}
    \includegraphics[width=0.49\textwidth]{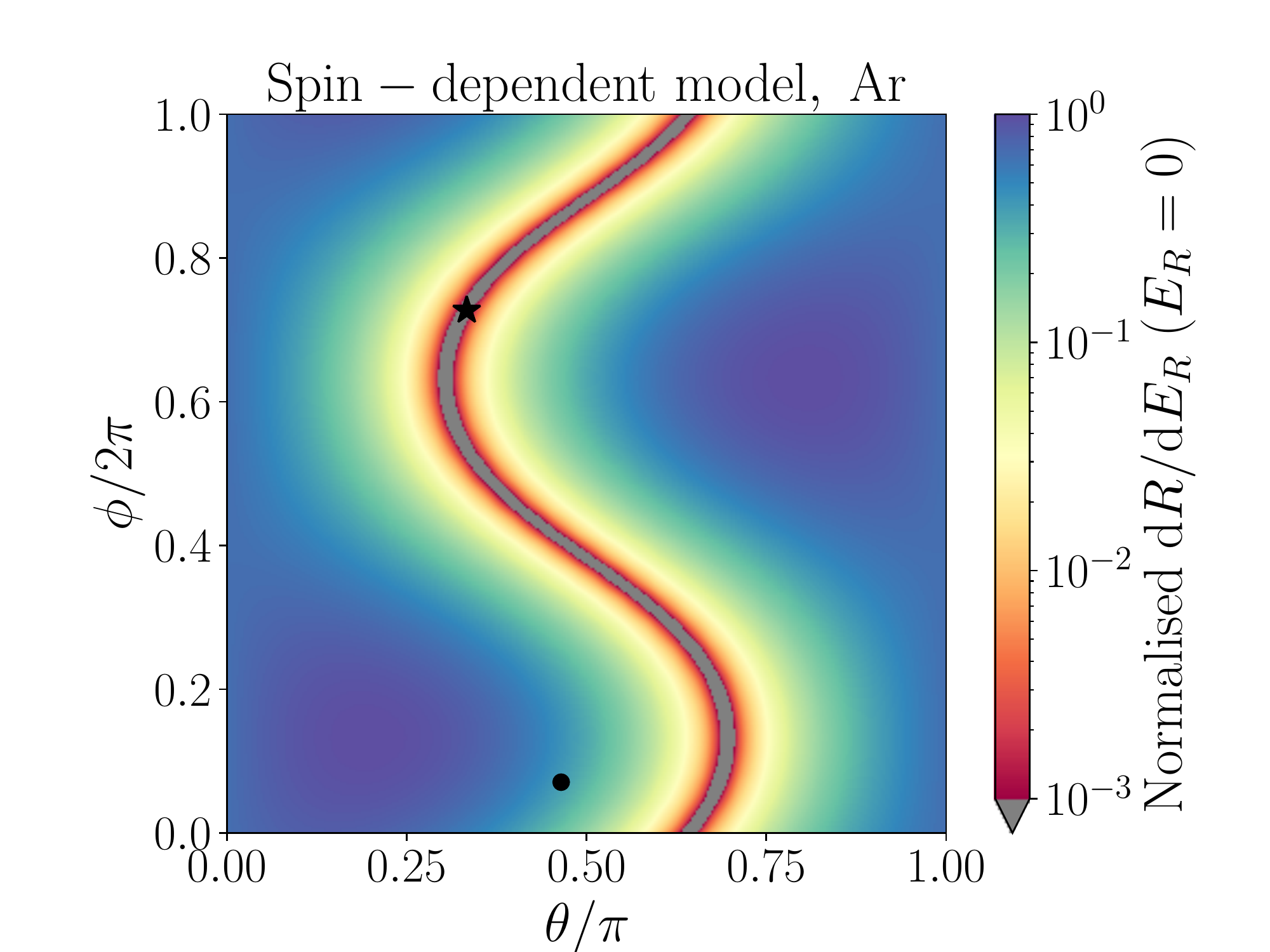}
    \caption{Relative ${\rm d}R/{\rm d}E_R$ for xenon (left) and argon (right) following the parametrisation in Eq.\,\eqref{eq:SIparam} for 100~GeV dark matter mass. The top row reflects results for the spin-independent vector-like couplings model (where the contribution of the $\mathcal{O}_{1}$, $\mathcal{O}_{7}$, and $\mathcal{O}_{9}$ operators is considered) while the bottom row represents the situation in the spin-dependent axial vector-like couplings model (with operators $\mathcal{O}_{4}$, $\mathcal{O}_{6}$,  $\mathcal{O}_{8}$, and $\mathcal{O}_{9}$). In the spin-independent case the contour of maximal suppression corresponding to a fixed neutron-to-proton coupling ratio is highlighted by a dashed line.
    The star and circle markers correspond to specific model points studied and detailed in Section~\ref{sec:results}.
    }
    \label{fig:difrate_param}
\end{figure*}

In the top row of Figure~\ref{fig:difrate_param} we show the relative size of the differential recoil rate Eq.\,\eqref{eq:difrate} for a xenon and argon target respectively. The rate depicted by the colour map is normalized simply to the highest value in the map itself. We do so according to the parametrisation in Eq.\,\eqref{eq:SIparam} and take the $E_R\rightarrow 0$ limit. It can be observed that there are substantial regions where there is a large suppression $\lesssim 10^{-1}$ of the differential rate. Furthermore, when one compares the suppression plot for different target nuclei, i.e.\ xenon and argon, the pattern is similar. This is because where the suppression occurs is largely dominated by $\mathcal{O}_1$ and is extremal when $c_n/c_p\approx -Z/(A-Z)$, where $A$ and $Z$ are the use atomic mass and proton numbers respectively. Note that the minimum of the colour map is not the true minimum, but is instead capped at $10^{-3}$ for aesthetic reasons.

Similarly, for the dark axial vector interaction, the prescription above is followed but replace $(\bar{\chi}\gamma_\mu\chi)\rightarrow(\bar{\chi}\gamma_\mu\gamma_5\chi)$, and $\left(Q_{2,i}^{(6)},Q_{3,i}^{(6)}, Q_{4,i}^{(6)}\right) \rightarrow \left(Q_{6,i}^{(6)},Q_{7,i}^{(6)}, Q_{8,i}^{(6)}\right)$ following the nomenclature of Ref.\,\cite{Bishara:2018vix}. As before, and exactly as in Ref.~\cite{Alanne:2022eem}, we can eliminate any redundancies by parameterizing the coefficients to these operators

\begin{eqnarray}
    C_{6,1}^{(6)}\rightarrow \frac{g^\prime}{\Lambda^2} &\cos{\theta}&,\,\,\,\, C_{7,1}^{(6)}\rightarrow \frac{g^\prime}{\Lambda^2} \sin{\theta}\cos{\phi},\nonumber\\
    &C_{8,1}^{(6)}&\rightarrow \frac{g^\prime}{\Lambda^2} \sin{\theta}\sin{\phi}.
    \label{eq:SDparam}
\end{eqnarray}

Performing the NREFT mapping, one ends up with non-zero contributions for $\mathcal{O}_{4}$, $\mathcal{O}_{6}$, $\mathcal{O}_{8}$, and $\mathcal{O}_{9}$. Similarly to the top row of Figure~\ref{fig:difrate_param}, we map out the suppression of the SD interactions within this parametrisation in the bottom row. Here the situation is substantially different, the suppression pattern for xenon and argon no longer resemble each other. For xenon, the condition of gauge invariant interactions provokes other unavoidable nuclear responses that diminish the suppression in many cases, as shown in Ref.\,\cite{Alanne:2022eem}. 
This is because the event rate associated with the various contributing operators are similar in size, unlike in the SI case. For an argon target, the pattern we see is much simpler because in the argon target only the $^{40}$Ar isotope is present in large quantities and it does not yield any response to the $\mathcal{O}_4,\,\mathcal{O}_6$, or $\mathcal{O}_9$ operators. The corresponding patterns then show a high degree of complementarity between the two targets for the SD model. Since the plots of Figure~\ref{fig:difrate_param} are normalised differential rates, the actual complementarity will depend on how sensitive xenon and argon target experiments will be to the unnormalised differential rate. The main point of showing Figure~\ref{fig:difrate_param} is to exhibit where suppression occurs in the gauge invariant parametrisation, in a way that is as detector independent as possible. In later sections, we show how these parameter choices affect interpretations of current and future experimental limits.
\begin{figure*}[htbp]
    \centering
    \includegraphics[width=0.99\textwidth]{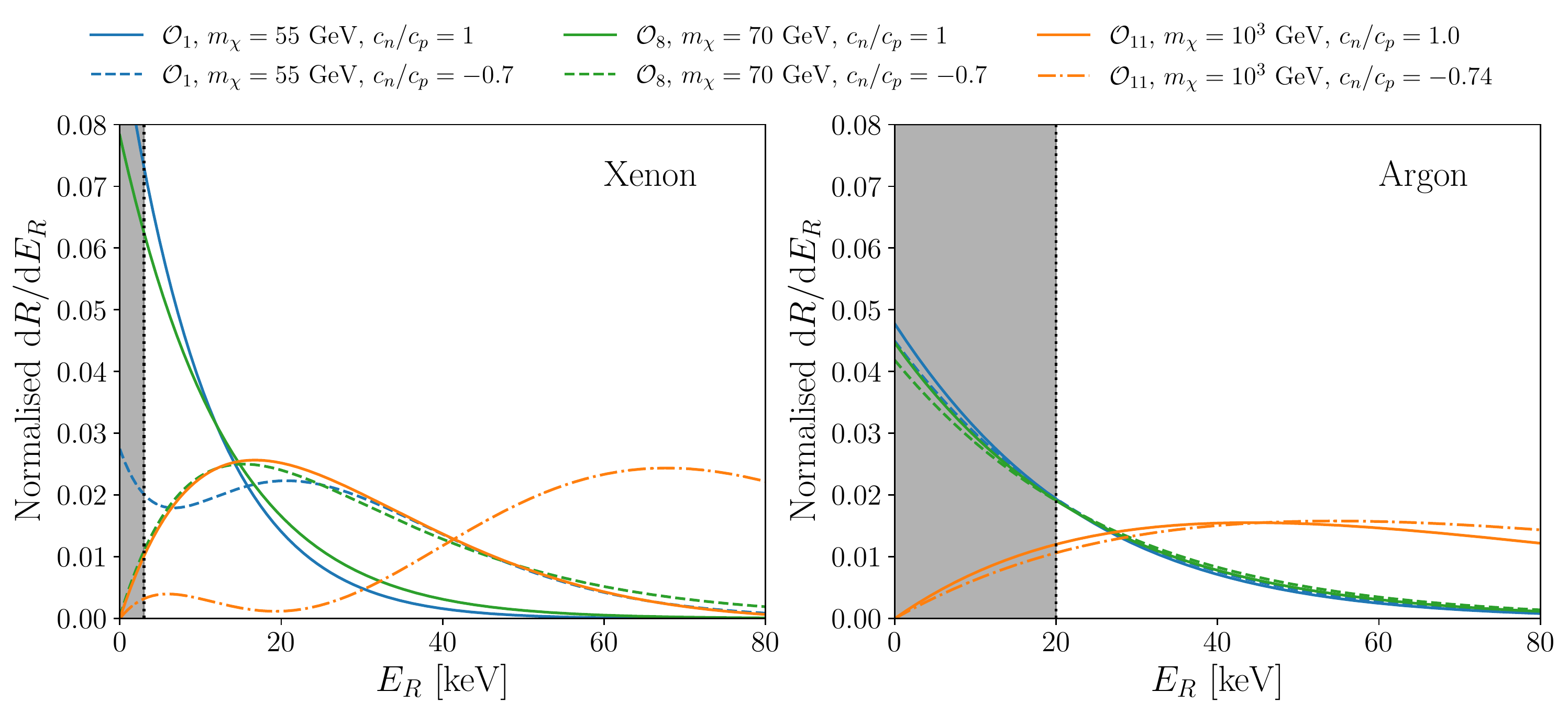}
    \caption{Recoil spectra in xenon (left) and argon (right) for varying interaction types, dark matter masses and neutron-to-proton coupling ratios, illustrating how isospin-violating interactions can induce large spectral changes in addition to overall rate. The grey-shaded region highlights regions typically inaccessible in the experimental setups considered here.}
    \label{fig:isospin_effect_spectra}
\end{figure*}

Furthermore, in order to minimize the experimental input of Figure~\ref{fig:difrate_param} we have chosen to show the differential recoil rate in the limit $E_R\rightarrow 0$. This means that $q$-dependent operators such as $\mathcal{O}_9$ do not affect the patterns shown. It should be noted that other works that investigated isospin violation simply took the suppression in the $E_R\rightarrow 0$ limit and use such suppressions to recast calculated limits~\cite{Frandsen:2011cg,Gao:2011ka,Drozd:2015gda,Lozano:2015vlv, Yaguna:2016bga,Li:2019sty}. Of course the presence of $q$-dependent operators means that this limit can no longer be taken. Additionally, doing this overlooks some interesting features of isospin-violating signals. To our knowledge, the most careful treatment to date is found in Refs.\,\cite{Liu:2017kmx, Brenner:2022qku} where the authors do consider $q$ and $v$-dependent operators and explore isospin violation in a model independent way. Ref.~\cite{Liu:2017kmx} specifically looks at the $c_n/c_p$ values that maximally suppress direct detection signal and how it varies with $m_{\chi}$, while Ref.~\cite{Brenner:2022qku} employs a statistical approach to derive limits that account for isospin violation. This article builds on the existing literature by showing how the extent of the suppression varies with $m_{\chi}$ as well as providing useful reference values for anyone with an isospin-violating model, our work also explicitly shows the important role argon-based detectors will have in this regard.

Figure~\ref{fig:isospin_effect_spectra} shows multiple spectral shapes for different NREFT responses, for a xenon target (left panel) and argon target (right panel). Because of the different strengths of nuclear response to certain operators, we show the differential rate normalised such that the integral over ${\rm d}E_R$ is 1, as this better illustrates the spectral variations. We also show roughly where the expected experimental thresholds will be, $E_{\rm th}$, by plotting grey shaded regions for $E_R\leq E_{\rm th}$. Let us first examine why moving beyond the $q\rightarrow 0 $ limit, even for momentum-independent operators such $\mathcal{O}_{1}$ (blue lines), may be important. We have chosen to show the isospin conserving case (solid) along with a result that is close to the maximum suppression for xenon, $c_n/c_p=-0.7$ (dashed). 
What we want to emphasise is that the isospin-violating spectrum goes to zero at significantly higher $E_R$, even for the same values of $m_\chi$, and does not exhibit the characteristic exponential shape that would be expected for a typical $\mathcal{O}_1$ interaction. Also notice how the rate for the $c_n/c_p=-0.7$ case in xenon is larger below $E_{\rm th}$, underlining the potential importance that detector effects may have on the expected level of suppression. 

The fact that the $c_n/c_p=-0.7$ spectrum for $\mathcal{O}_1$ in the signal region is not monotonically decreasing is notable because this behaviour is not expected for $q$-independent recoils. We exhibit the behaviour of the momentum-dependent operator $\mathcal{O}_{11}$ with $c_n/c_p=1$ for comparison. We have deliberately chosen a value of $m_{\chi}$ such that the spectra start to resemble each other. Exemplifying that this can also occur in velocity-dependent operators, we include the $\mathcal{O}_8$ spectrum, which much like $\mathcal{O}_1$, is monotonically decreasing in the isospin conserving case. However, when $c_n/c_p=-0.7$ we see that the spectrum matches very well with $\mathcal{O}_{11}$, even when the values for $m_{\chi}$ are really very different. We have additionally included the $c_n/c_p=-0.74$ with $\mathcal{O}_{11}$ case because we believe that this spectral shape is quite unique. Potentially there are inelastic dark matter models that would replicate this feature. 

Other than underlining the importance of doing spectral analysis when determining how much suppression one expects, Figure~\ref{fig:isospin_effect_spectra} shows how, in the event of signal detection, specific isospin violating configurations may lead to signal misidentification. The right panel in Figure~\ref{fig:isospin_effect_spectra} shows the exact same benchmarks, but now the recoil spectra is with an argon target. We see that the effect on the signal shape is rather small. This is because the suppression for argon peaks at a different value of $c_n/c_p$. 
Post-discovery, once the overall rate of an observation is also taken into account, correct operator identification should be possible with both a xenon and argon target experiment with comparable and requisite sensitivities. This stresses the importance of complementarity between xenon and argon detectors that is present even in the SI particle model. 

The non-standard signal shapes in Figure~\ref{fig:isospin_effect_spectra} come directly from the nuclear form factors and how they interfere at different recoil energies, see Eq.\,\eqref{eq:diff_crosssec}. Therefore, the consequences of isospin-violating models are highly dependent on nuclear responses. In order to check that our results will be in some way robust to any uncertainty in these nuclear form factors, we would require a systematic way of incorporating them into our analysis. However, as far as we are aware, this has not been made available, at least for the NREFT basis, see Ref.\,\cite{Cerdeno:2012ix} for a discussion of SD structure functions. We leave a systematic study to future work, and instead present here illustrative examples of how, in the regions of parameter space where rate suppression from isospin violation is important, the choice of nuclear form factor has a large impact on the expected signal. 

\begin{figure*}[htbp]
    \centering
    \includegraphics[width=0.99\textwidth]{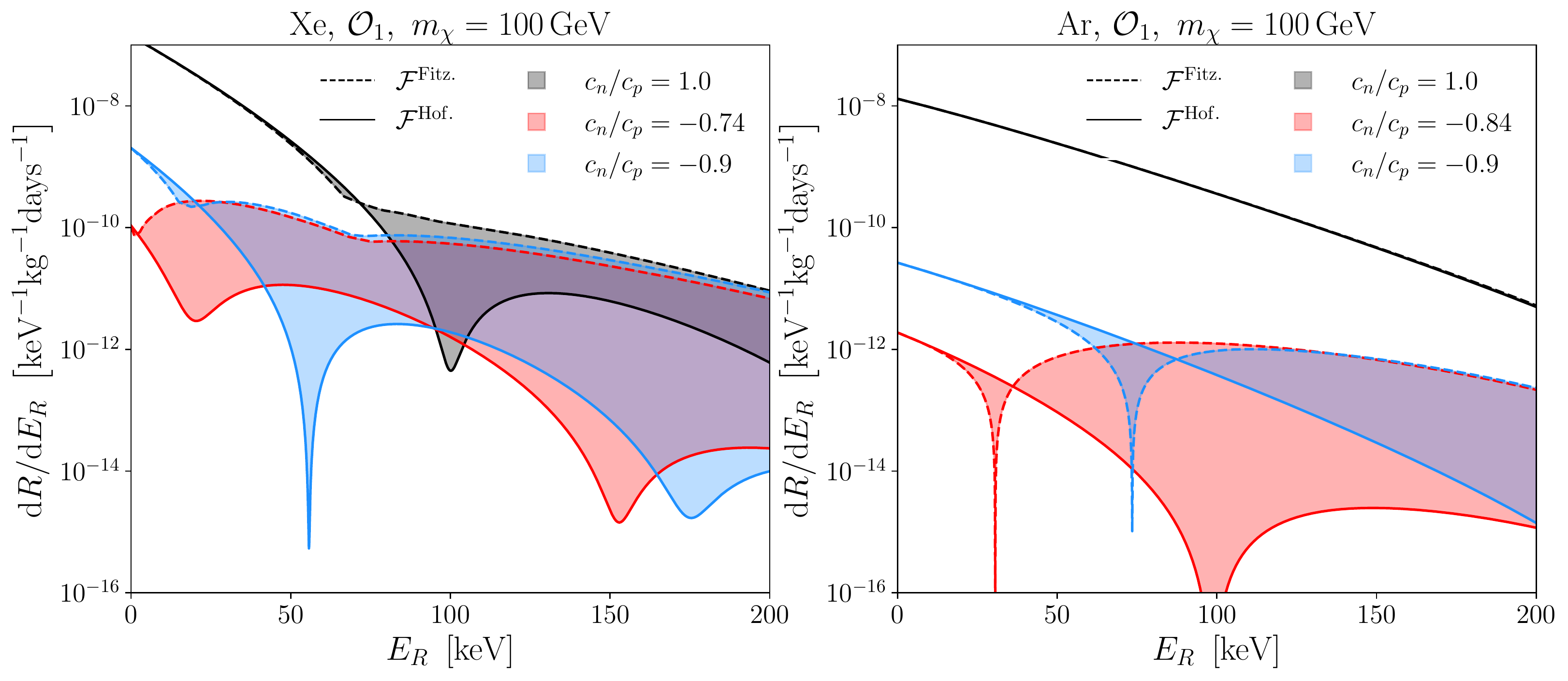}
    \caption{Difference in the recoil spectra in xenon (left) and argon (right) for dark matter mass of 100~GeV between the two sets of form factors (dashed versus solid lines), for three different coupling ratios. Note we have set $c^p_1 \langle v\rangle_{\rm Higgs}=10^{-5}$ and $c^n_1=c_n/c_p \times c^p_1$ where $c_n/c_p$ is the isospin ratio as signified in the caption. }
    \label{fig:FormFactor_difference}
\end{figure*}

In Figure~\ref{fig:FormFactor_difference}, we compare the $\mathcal{O}_1$ form factors from Refs.\,\cite{Fitzpatrick:2012ix, Hoferichter:2018acd}, which we name $\mathcal{F}^{\rm Fitz.}$ and $\mathcal{F}^{\rm Hof.}$ after the first authors of each reference respectively. We are unaware of any reference where such a large difference has been pointed out. Recent works that use the $\mathcal{F}^{\rm Fitz.}$ form factors are \cite{Rogers:2016jrx, Bozorgnia:2018jep, Cerdeno:2019vpd,  LUX:2020oan, SuperCDMS:2022crd, Cappiello:2022exa} whereas works like Ref.\,\cite{XENON:2022avm} use $\mathcal{F}^{\rm Hof.}$. Other works such as Refs.\,\cite{Gorton:2022eed,Kang:2022zqv} and popular codes such as DDcalc~\cite{Athron:2018hpc}, WIMPy~\cite{WIMpy-code} and WimPyDD~\cite{Jeong:2021bpl} use Ref.\,\cite{Anand:2013yka}, which we find much more closely resembles that of $\mathcal{F}^{\rm Hof.}$ than $\mathcal{F}^{\rm Fitz.}$.

Figure~\ref{fig:FormFactor_difference} shows the differential recoil rates for the $\mathcal{O}_{1}$ interaction for xenon (left) and argon (right) targets. We see that in the zero momentum limit, the form factors are in agreement. At higher $E_R$, they start to disagree, underlining the importance of both having a reliable understanding of nuclear responses and determining the isospin suppression within the signal region and not just in the $E_R\rightarrow 0$ limit We can see that the differential rates for a xenon target are quite sensitive to the choice in form factor. The shaded region, highlighting the difference between $\mathcal{F}^{\rm Fitz.}$ and $\mathcal{F}^{\rm Hof.}$, can be very large, even when the couplings are not so close to the maximal suppression of $c_n/c_p=-0.9$. Notice as well that the $\mathcal{F}^{\rm Hof.}$ form factors tend to be more suppressed over the whole range than $\mathcal{F}^{\rm Fitz.}$ for xenon. For argon, the simpler nucleus, the variation in rate is negligible in the isospin-conserving case. Additionally, when the variations are large as with the $c_n/c_p=-0.84$ and $c_n/c_p=-0.9$, we see that the effects tend to start at higher $E_R$. In Appendix~\ref{app:formfact_suppression} we show how isospin suppression is affected by the choice of form factors and the effect can be substantial. In Appendix~\ref{app:formfact_suppression}, we observe that the largest difference between form factors occurs for $\mathcal{O}_1$ when isospin suppression is most pronounced. This is because the behaviour in the limit as $E_R\rightarrow 0$ no longer dominates over the signal. Moreover, since isospin suppression involves a fairly fine cancellation between form factors, and over a range of $E_R$, it makes sense that the regions of greatest suppression are particularly sensitive to changes in the way form factors are parameterised.

It is beyond the scope of this paper to speculate on the origin of these differences, but given the size of the effect, we wanted to point it out. We are aware that considerable work is being undertaken currently to independently verify the accuracy of the nuclear form factors and to determine any sources of theoretical uncertainty\,\cite{Hoferichter:2018acd,Hoferichter:2019uwa,Khaleq:2022mwr}. Furthermore, we see this as further motivation for the state-of-the-art techniques that accurately model nuclear responses as well as encouraging a further investigation of the uncertainties associated with such calculations. 

In terms of accuracy, we believe that Refs.\,\cite{Hoferichter:2018acd,Hoferichter:2019uwa} are the current gold standard for SI interactions, and we will continue with our study using these. We take this position because Ref.\,\cite{Hoferichter:2018acd} states that they have made use of improvements in nuclear modelling, including multi-nucleon interaction effects, to provide more precise form factors. Additionally, authors of Ref.~\cite{Hoferichter:2018acd} go to much greater lengths to validate their nuclear structure calculations by comparing nuclear spectra of their shell model to experimental results. For the SD responses, we use the form factors in Ref.\,\cite{Anand:2013yka}. 

\section{Experimental setup and limit evaluation}
\label{sec:exp_setup}

In this section we outline our derivation of bounds and projections for the direct detection experiments we consider. As mentioned above, we focus on xenon and argon targets, which are the basis for what are or will be the most sensitive experiments for $m_{\chi}\gtrsim 10$ GeV when one considers S1 (scintillation) and S2 (ionisation) analyses. Currently, the most recent argon experiments, DarkSide-50~\cite{DarkSide:2018kuk} and DEAP-3600~\cite{DEAP:2019yzn} are substantially less sensitive than XENON1T~\cite{Aprile:2018dbl} due to either differences in exposure ($0.046$~tn~yr for DarkSide-50 versus 1~tn~yr for XENON1T), or due to efficiency in the case of DEAP-3600. However, the argon detector that will be taking data in this decade, DarkSide-20k~\cite{Aalseth:2017fik}, with its projected 200~tn~yr exposure, will be much more competitive with LZ~\cite{Mount:2017qzi} and XENON-nT~\cite{XENON:2020kmp} that are expected to have 15.33 and 20~tn~yr exposures respectively. It is this set of experiments in which we are basing our projections for future-Xe and future-Ar experiments. We do not attempt to project sensitivities for next-generation experiments similar to that of ARGO~\cite{ARGO_Snowmass} or DARWIN/XLZD~\cite{DARWIN:2016hyl} because they are at an earlier stage of their development and many experimental parameters and characteristics are yet to be determined or finalised.

We perform our analysis in terms of recoil energy and do not simulate detector responses in terms of S1 and S2 signals. We, therefore, do not replicate the full S1+S2 two-dimensional likelihood calculation performed by the experimental collaborations. To obtain the number of expected recoil events, $N$, from the differential rate, Eq.\,\eqref{eq:difrate}, one needs to include detector specific information, such as exposure, $\varepsilon$, the energy-dependent acceptance $\epsilon(E_R)$ and detector resolution. In particular, the resolution is modeled as a Gaussian and is accounted for via a convolution, 
\begin{equation}
    \label{eq:n_dd}
    N = \varepsilon\int_{E_\mathrm{th}}^{E_\mathrm{max}} \left(\int_0^\infty \frac{\dd R}{\dd E^\prime}\epsilon(E^\prime)\frac{e^{-\frac{\left(E_R - E^\prime\right)^2}{2\sigma^2(E^\prime)}}}{\sigma(E^\prime)\sqrt{2\pi}}\,\dd E^\prime\right) \dd E_R\,.
\end{equation}
Additionally, one might need to include an acceptance cut, which typically is some multiplicative factor and is sometimes already incorporated in the efficiency. 

For XENON1T we make use of the public code associated with Ref.~\cite{XENON:2022zkh} to carry out an exclusion limit calculation in terms of the reconstructed energy, which approximates the XENON1T likelihood very well. This allows us to input a recoil spectra in energy units and we do not need to apply any additional acceptance, efficiencies or resolution in addition to what is being calculated in the package.
For the other experiments that we consider in this work, the efficiencies are readily available in the literature. For our future xenon detector, we use the efficiency curve of LZ from Ref.\,\cite{LZ:2022ufs} which already has the acceptance incorporated. For DarkSide-50, the curve from Figure~10 of Ref.\,\cite{DarkSide:2018kuk} is used. For DEAP-3600 we use the information from Figure~21 of Ref.\,\cite{DEAP:2019yzn}. For our future argon sensitivity projections, we use the DarkSide-20k NR acceptance, taken from Figure 92 of Ref.\,\cite{Aalseth:2017fik}.

In order to model resolution effects we take the detector-dependent width $\sigma(E^\prime)$, with $E^\prime$ being the ``true'' recoil energy. For XENON1T we use the function in Ref.\,\cite{XENON:2020rca}, whereas for our future xenon experiment we use the function given in Ref.\,\cite{LUX:2016rfb}. For DarkSide-50, DEAP-3600, and our future argon detector we use the resolution function given in Ref.\,\cite{Pagani:2017wuk}. The resolutions are  given as functions of electron equivalent recoil energy, $E_{\rm ee}$. In order to translate from nuclear recoil $E_R$, we make use of modified Lindhard factors~\cite{osti_4701226,Sorensen:2011bd}. For xenon detectors we use the numerical fit for the parameter $k$ in the Lindhard model determined in Ref.\,\cite{LUX:2017bef}, whereas for argon, we make use of the effective Lindhard factor experimentally determined in Ref.\,\cite{Szydagis:2021hfh}. We have assumed that the electronic recoil receives negligible amounts of quenching allowing us to interpret $E^\prime$ in as the electron equivalent energy $E_{ee}$. We note that this assumption may not be entirely true and for different experiments, the precise behaviour of ionised electrons and scintillation photons and how their energy is divided may alter how one models the resolution. The method implemented in this work is sufficient for this study and a more sophisticated analysis, like those performed in experimental collaborations will not change the results of this paper.

In order to calculate an exclusion limit or projected sensitivity of an experiment, one needs to determine the expected background signal, $N_{\rm b}$ as well as the observed number of events, $N_{\rm obs}$. We use the Poissonian probability, 
\begin{equation}
\mathcal{P}\left(N_{\mathrm{obs}} | N_{\mathrm{th}}\right)=\frac{N_{\mathrm{th}}^{N_{\mathrm{obs}}} e^{-N_{\mathrm{th}}}}{N_{\mathrm{obs}} !}
\end{equation}
where $N_{\rm th}=N_{\mathrm{DM}}+N_{\mathrm{b}}$ is the sum of the expected background plus dark matter signal given by Eq.~\eqref{eq:n_dd}. 
For both DarkSide-50 and DEAP-3600, no background events were expected and no recoils were observed. 

For our projections for DarkSide-20k and LZ, we use binned spectra for the expected background and signal rather than a total counts. The Likelihood from $k$ independent energy bins is simply the product of the Poissonian probability of each bin, and we build a test statistic (TS) from the Likelihood,

\begin{equation}
 \mathrm{TS}(\lambda)=-2 \log \left(\frac{\operatorname{Likelihood}\left(N_{\mathrm{obs}} | N_{\rm{b}}, N_{\rm{DM}}\right)}{\operatorname{Likelihood}\left(N_{\mathrm{obs}} | N_{\mathrm{b}}\right)}\right).
    \end{equation}
such that, using the Poisson probability, the TS can be written as 
\begin{equation}
\mathrm{TS}(\lambda)=\sum_{k}\left[-2 N_{\mathrm{obs}, k} \log \left(\frac{N_{\mathrm{DM}, k}+N_{\mathrm{b}, k}}{N_{\mathrm{b}, k}}\right)+2 N_{\mathrm{DM}, k}\right]
\end{equation}
where the sum over $k$ is a sum over the bins. This test statistic can be used to determine a two-sided confidence interval.
In this case in order to find the 90\% exclusion limit or projected sensitivity, the TS is approximated as a $\chi^2$ distribution.

With our future xenon and argon experiment projections being inspired by LZ and DarkSide-20k respectively, we consult the backgrounds reported by the relevant collaboration. 
For the DarkSide-20k inspired argon detector, the background discrimination between ER and NR events using pulse shape discrimination is expected to be extremely effective, resulting in a near background-free experiment (to a level of $<0.1$ events in the entire exposure~\cite{Aalseth:2017fik}). After ER backgrounds have been discriminated the only background remaining is the CE$\nu$NS background from solar neutrinos~\cite{BATTISTONI2005526,Billard:2013qya}, expected to contribute $\sim 1.6$ background events within a 100 tonne-year exposure. We use background and signal spectra in a binned distribution with 1~keV width bins between 20 and 200~keV in terms of the nuclear recoil energy.

In order to attempt to replicate the LZ first search results with an exposure of 60 live days, we use the background spectrum from Figure~6 from Ref.\,\cite{LZ:2022ufs} in units of the electron equivalent energy with 3 bins per keV$_{ee}$ between 0 and 17~keV. The dark matter spectra are calculating using the same binning, using Lindhard factor to convert between $E_R$ and $E_{ee}$. 
To reproduce the effect of the S1--S2 background discrimination employed in the LZ collaboration's first data analysis we make use of Figure~4 from Ref.\,\cite{LZ:2022ufs} and scale down the backgrounds to total 11 events (compared to 333 originally in the background distribution). As this region corresponds to the NR 90\% quantile region, we also apply a 0.9 flat acceptance to the dark matter signal. This allows us to replicate the expected limit well for the entire mass range, and the observed limit at higher dark matter masses. However, we cannot perfectly match the observed limit in the region of maximum sensitivity around $20-30$~GeV dark matter mass, which is due to an under-fluctuation of observed events in the S1--S2 search space, due to lack of public information on the exact distribution of expected backgrounds. 
The official LZ exclusion limit at around $20-30$~GeV is around three times more stringent than that of our calculation. 
Our ability to replicate the LZ results is consistent in performance to the updated treatment of the LZ limits as implemented in DDcalc~\cite{GAMBITDarkMatterWorkgroup:2017fax,GAMBIT:2018eea} by the GAMBIT working group~\cite{Chang:2022jgo}.
For the XENON1T limit, using the approximate likelihood code gives good agreement with the published limit.

For the argon experiments, we see a very good match with DarkSide-50 and DEAP-3600 for all dark matter candidate masses. For the DarkSide-20k projection, we can match the projection from the 2017 TDR very well by using a 1-bin method with 1.6 neutrino background events per 100 tonne-years of exposure. However, in this paper we use a binned spectrum which gives us an additional 20--50\% sensitivity for most masses. Using binned spectra rather than a total number of counts allows us to determine the effect of the spectral shape differences in the isospin-violating models as well as the difference in total rate. Our analysis is able to replicate the isoscalar ($c_n/c_p=1$), isovector ($c_n/c_p=-1$) and xenophobic ($c_n/c_p=-0.7$) working point limits for five different NREFT operators presented by DEAP-3600 in Ref.\,\cite{DEAP:2020iwi}.

For our projections for the whole 1000 live days exposure of LZ we follow the same method as described above used for the observed limit with 60 live days, but with backgrounds scaled up for the exposure and assume the observed spectra equals the expected backgrounds.

We note that some of the above methods described are applied on data sets with low expected backgrounds. In these cases Wilks' theorem may no longer hold and the test statistic no longer converges to $\chi^2$, meaning the drawing of limits may be affected \cite{Baxter:2021pqo}. We have checked that isospin suppression affects both of the experimental limits calculated with the methods described above and using toyMCs in the same way. Since our simplified limits replicate better the results from the full experimental analyses, we present our results using these.

\section{Results}
\label{sec:results}
In this section we show the results for isospin-violating scenarios, using the detector setup described above. First, we present the work for the two gauge invariant cases described in Section~\ref{sec:NREFT_GImodels}. Second, we present results considering only one NREFT operator at a time. We do this for an incomplete set of operators so as to not overload the main text but results for other operators can be found in Appendix~\ref{app:2D}.  

\subsection{Spin-(in)dependent model}
\label{subsec:results_GI}

\begin{figure*}[htbp]
    \centering
    \includegraphics[width=\textwidth]{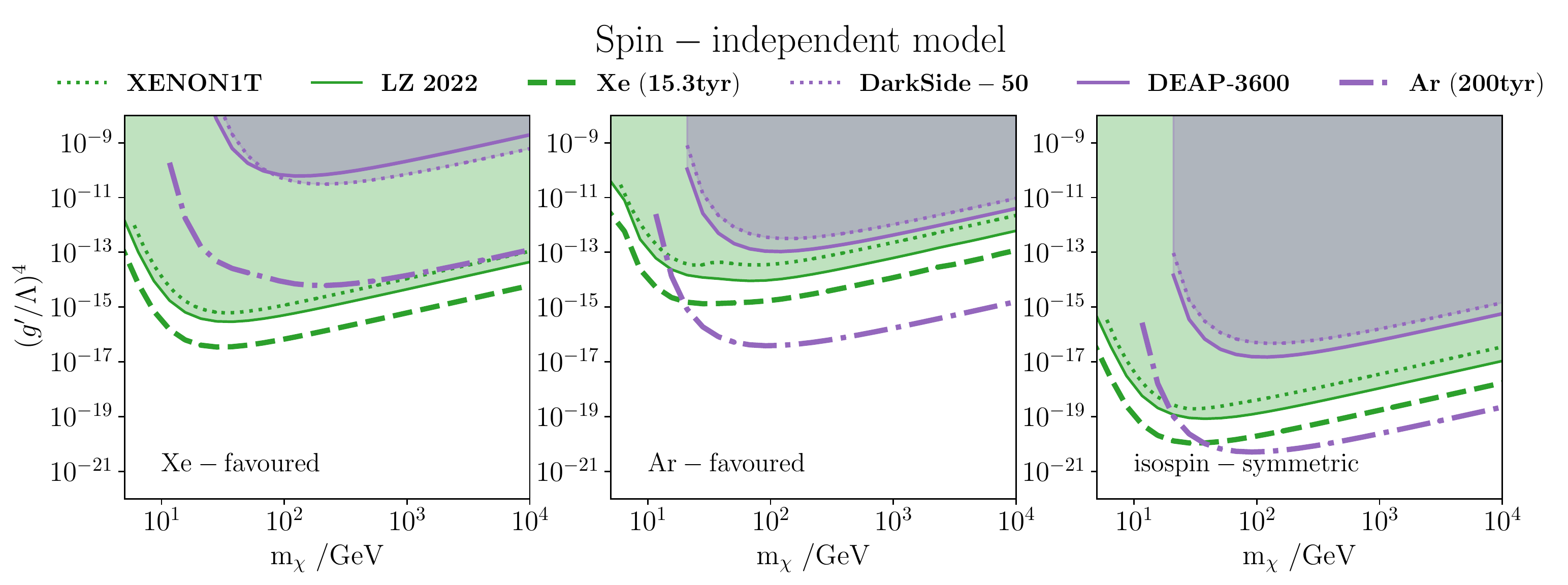}\\
    \includegraphics[width=\textwidth]{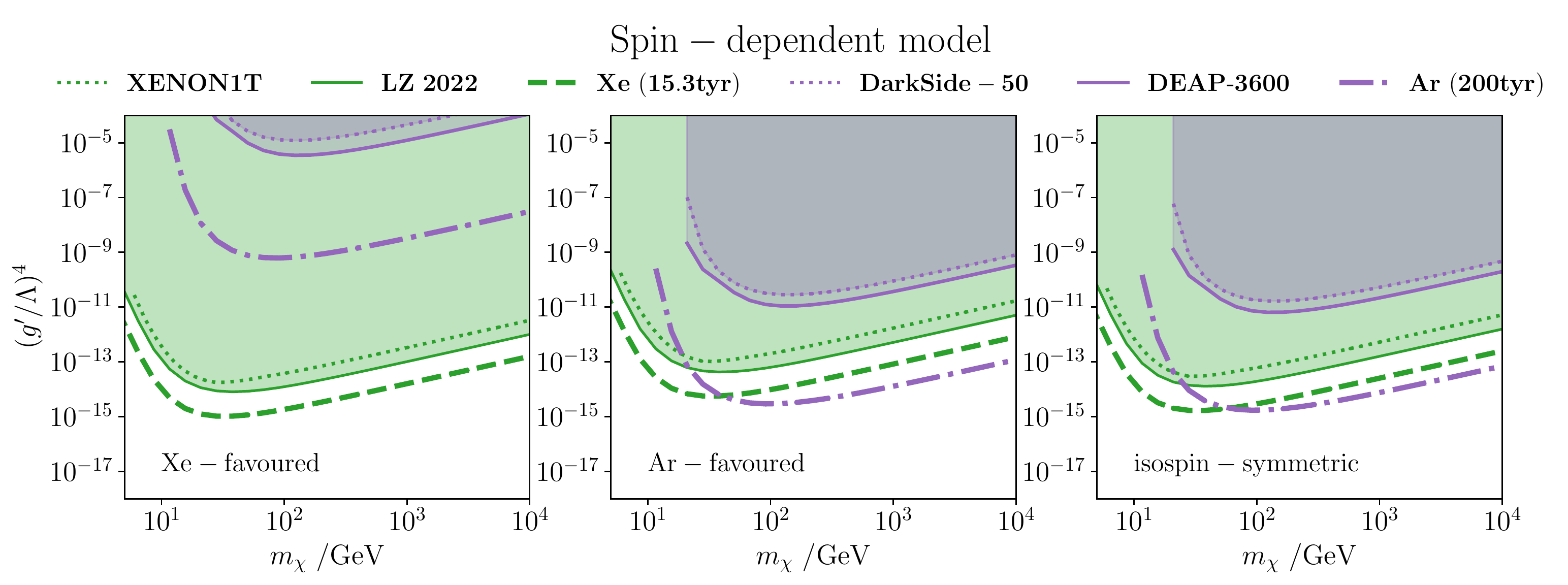}
    \caption{Current LXe and LAr dark matter exclusion regions [shaded] and future projections for two near-future liquid noble direct detection experiment sensitivities (a 200 tonne-year argon experiment and a 15 tonne-year xenon experiment) [dashed]. \textbf{Top:} spin-independent model incorporating the $\mathcal{O}_1$, $\mathcal{O}_7$, and $\mathcal{O}_9$ operators. \textbf{Bottom:} spin-dependent model incorporating the $\mathcal{O}_4$, $\mathcal{O}_6$, $\mathcal{O}_8$, and $\mathcal{O}_9$ operators. We show different isospin configurations, a xenon-favoured scenario (left), an argon-favoured scenario (middle), and the isospin-conserving case (right).
    }
    \label{fig:sens-gauge-inv}
\end{figure*}

As described in Section~\ref{sec:NREFT_GImodels}, a gauge invariant formulation of dark matter interactions leads to the requirement of additional operator responses beyond the standard $\mathcal{O}_1$ or $\mathcal{O}_4$ formalism. For the SI model, this is of little consequence because of the large sensitivity difference between $\mathcal{O}_1$ compared to $\mathcal{O}_{7}$ and $\mathcal{O}_{9}$. Nonetheless, our calculations of the leading xenon and argon-based direct detection experiments as well as the future projections serve as a timely update to the literature. This is even more true given the form factor dependence shown in Figure~\ref{fig:FormFactor_difference}. The top row of Figure~\ref{fig:sens-gauge-inv} shows the current and future sensitivities for the SI model. The panels show isospin configurations that are phenomenologically interesting, from left to right we have xenon-favoured, argon-favoured, and isospin-conserving scenarios. We choose the Xe/Ar-favoured parameter values by taking the ratio of total events in the LZ exposure and the DarkSide-20k exposure in the $\phi$ and $\theta$ parameter space, and finding the minimum and maximum points. 
For the spin-independent model this method results in an argon-favoured point of $\theta/\pi=0.62$ and $\phi/{2\pi}=0.98$, however, there is a range of values with approximately the same relative argon-to-xenon suppression ratio that we could have chosen, these follow the $c_n/c_p=-0.72$ contour shown in Figure~\ref{fig:difrate_param}. The equivalent maximal xenon-favoured point is found at values of $\theta/\pi=0.45$ and $\phi/{2\pi}=0.41$. These Ar-favoured and Xe-favoured points have been highlighted on Figure~\ref{fig:difrate_param} using circle and star markers, respectively.

In the SD case, the situation is more interesting because gauge invariance requires that operator $\mathcal{O}_{4}$ is accompanied by $\mathcal{O}_{6}$, $\mathcal{O}_{8}$, and $\mathcal{O}_{9}$. Here the different operators provoke nuclear responses that have roughly the same normalisations. From Figure~\ref{fig:difrate_param}, we saw that there should be a high degree of complementarity between argon and xenon target searches. The bottom row of Figure~\ref{fig:sens-gauge-inv} shows the current and future sensitivities for the SD model. Once again we show three panels showing the xenon-favoured, argon-favoured and isospin-conserving scenarios. 
In the SD model, the relative dependence of the argon and xenon suppression rates with $\theta$ and $\phi$ is more complex due to the contributions of various xenon-sensitive operators. 
This results in the argon-favoured configuration not coinciding with the region of maximum xenon suppression (which can be seen in the bottom left of Figure~\ref{fig:difrate_param}). This is because where maximal xenon suppression occurs, the argon rate is similarly suppressed. Instead we find the argon favoured region occurs at a value of $\theta/\pi=0.46$ and $\phi/{2\pi}=0.07$.
The xenon-favoured configuration does however align with the region of maximum argon suppression (albeit not the region of minimum xenon suppression) at $\theta/\pi=0.33$ and $\phi/{2\pi}=0.72$, again demonstrated by the star marker in the bottom panel of Figure~\ref{fig:difrate_param}. For the isospin conserving point we use a value of $\phi=\pi/4$, for which the proton coupling equals the neutron coupling for all operators involved, for both models. At this $\phi$ value, isospin conservation holds for any value of $\theta$ and we show the limits for $\theta=0$.

Figure \ref{fig:sens-gauge-inv} shows the experimental exclusion limits which were calculated using existing public data from DarkSide-50, DEAP-3600, XENON1T, and LZ (with 60 days livetime), and derive projected sensitivities for DarkSide-20k and LZ (1,000 days livetime). For both the SI and SD models, results are presented as an exclusion limit at the 90\% confidence level on $(g^\prime/\Lambda)^4$, as this quantity is proportional to the dark matter event rate and therefore scales with exposure in the same way as familiar zero-momentum limit cross-section constraints.

We note that the effects of isospin violation can have dramatic impacts on the apparent parameter space of dark matter currently excluded and expected to be probed in the future. For the SI case, Figure~\ref{fig:sens-gauge-inv} shows a suppression at $m_{\chi}\sim 10^3\,{\rm GeV}$ of $\sim 10^6$ and $\sim 10^5$ for argon and xenon respectively. We see that currently, due to relatively small target exposures, argon limits are unable to lead in sensitivity even in the argon-favoured scenario, but this situation will change dramatically in the future. Future detectors will be of similar sensitivity, where at high $m_{\chi}$, the isospin-symmetric $\sim10$ difference is simply in line with the expectations due to the relative exposures. We see that this difference can be increased or decreased by $\sim 10^2$ or $\sim 10^{3}$ depending on the parameters chosen. Results of this kind have been widely reported in the literature before, but we feel having a dedicated up-to-date result will be instructive for the direct detection community. Furthermore, by including the effects of operators $\mathcal{O}_7$ and $\mathcal{O}_9$ we have been able to verify that even in manifestly gauge invariant models, the suppression scales seen by just considering $\mathcal{O}_1$, are not greatly affected.

Of greater novelty are the consequences we report for argon detectors in the gauge invariant SD model. The bottom row of Figure~\ref{fig:sens-gauge-inv} shows how argon detectors are currently sensitive to SD dark matter models and will be able to beat xenon detectors in the future. We do see across the three panels, that the xenon sensitivities are fairly stable, whereas for argon the sensitivity in the high $m_\chi$ region varies substantially, by a factor of $\sim 10^5$. 
This is thanks to the additional operators required by gauge invariance which substantially limits the points in parameter space where xenon suppression maximally occurs, see Fig.~\ref{fig:difrate_param}. Additionally, the regions where xenon is most suppressed are not aligned with areas where argon is least suppressed. If instead of showing the xenon and argon favoured points in Fig.~\ref{fig:1d-suppressions_SI}, we showed minimally and maximally suppressed xenon (argon) we would see a factor $\sim 10^3$ ($\sim 10^6$) difference in the respective limits.

We see that even in the isospin-conserving scenario, the future argon detector is more sensitive at high masses. This is quite the coup for a target that is insensitive to the canonical SD nuclear response, $\mathcal{O}_4$, and is simply a consequence of insisting on SM gauge invariance of the dark matter model under consideration. As shown in Ref.\,\cite{Alanne:2022eem}, SM gauge invariance requires that the $\mathcal{O}_8$ operator is nonzero, provoking a nuclear response from the argon atom. Since our Majorana fermion model is simply motivated by the dark axial vector current $\bar{\chi}\gamma^\mu\gamma^5\chi$, the standard source of spin-dependent interactions, we believe this result to be fairly general. 

In this paper we only explicitly consider data and experimental analyses from LXe/LAr TPCs using both the S1 and S2 ionisation/scintillation signals targeting $\gtrsim 10$~GeV mass dark matter candidates. However, we note that the XENON1T and DarkSide-50 experiments have established~\cite{XENON:2019gfn,XENON:2019zpr,DarkSide-50:2022qzh,DarkSide:2022dhx} that an ionisation-only analysis accounting for the impact of the Migdal effect and bremsstrahlung on electronic recoils allows for probes of dark matter to masses as low as 85~MeV and 40~MeV, respectively. Considering the published sensitivities of XENON1T and DarkSide-50 for dark matter masses below 10~GeV and the suppression results for the next generation experiments derived in this paper, we estimate that the relative dominance of future xenon and argon experiments illustrated in Figure~\ref{fig:sens-gauge-inv} at intermediate-to-high masses would extend down to the $\mathcal{O}(10)$~MeV-scale if data from ionisation-only analyses were also considered. An exception to this that we highlight is that it may be possible for DarkSide-20k to ultimately be more sensitive that LZ in the low mass region in the SI model even in the Xe-favoured regions of parameter space.

\subsection{NREFT results}
\label{subsec:results_NREFT}

In our view, the two models considered above are well-motivated and allow us to exhibit the main points of this paper. We are aware that there are numerous other possible candidates for dark matter and many of which are as theoretically attractive. In lieu of producing a comprehensive study of all the possible isospin-violating models, we do the next best thing and perform a study on NREFT operators in isolation. Those with their own dark matter models will be able to use the results presented here to estimate the implications isospin violation may have on the expected direct detection signal in current and future experiments.

\begin{figure*}[htb]
    \includegraphics[width=0.9\textwidth]{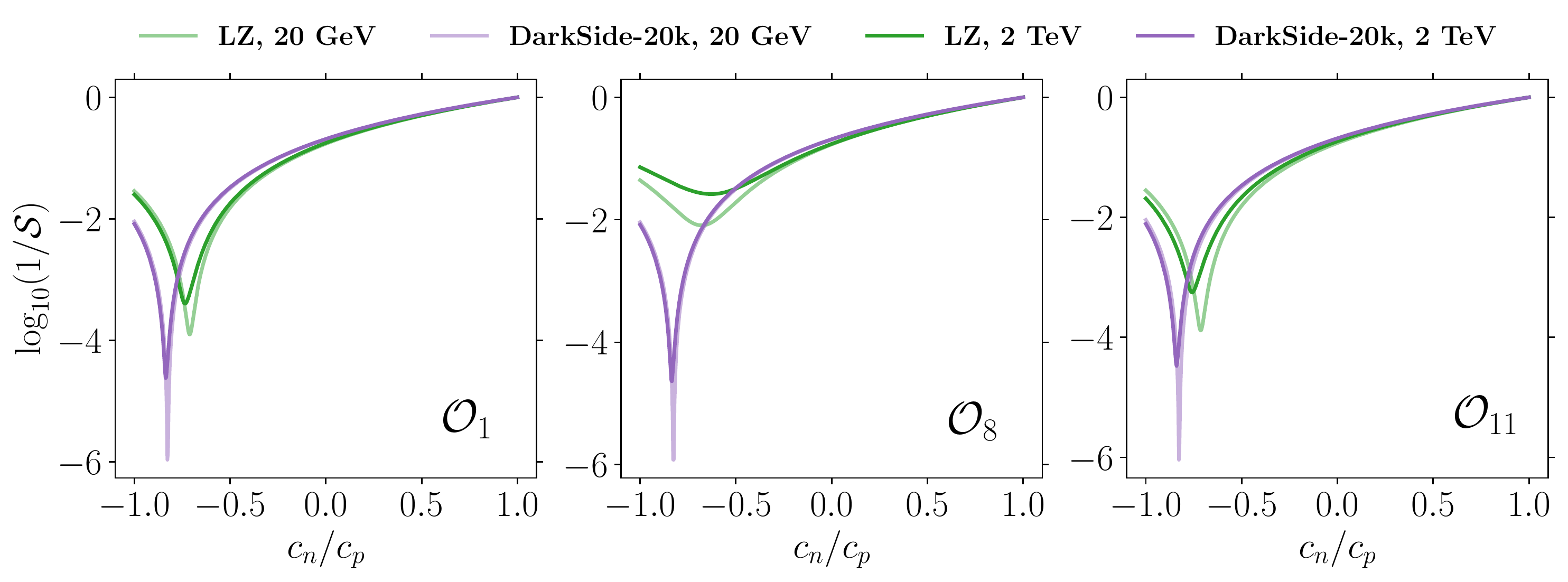}
    \caption{Inverse of the suppression factor for the experimental constraints on the dark matter -- nucleon interaction strength with neutron-to-proton coupling ratio, for a LZ-like and DarkSide-20k-like experiment. Shown here is a comparison of the limit suppression factors for EFT operators $\mathcal{O}_{1}$, $\mathcal{O}_{8}$, and $\mathcal{O}_{11}$ for dark matter masses 20~GeV and 2~TeV.
    }
    \label{fig:1d-suppressions_SI}
\end{figure*}

The way we present our results is quite different from the previous section. This is because we want to maximise the utility for interested parties and present results in a way that comprehensively covers the parameter space. Therefore, we want to convey how expected counts is affected over a range of $c_n/c_p$. Additionally, by performing our analysis in a more realistic way, i.e.\ not simply observing suppressions in the $E_R\rightarrow 0$ limit, we are now sensitive to variations of the signal shape coming from the nuclear response functions, as outlined in Section~\ref{sec:NREFT_GImodels}. This introduces a mass dependence on the suppression predicted because lower values of $m_{\chi}$ have their spectral shape dominated by the tail of the dark matter velocity distribution, and the light incident dark matter is unable to impart enough momentum transfer to probe the structure function. For larger $m_{\chi}$, the signal is shallower, meaning that the isospin cancellation has to persist over a greater energy region.

We quantify the impact of a change in coupling ratio on the dark matter -- nucleon coupling rate (or Wilson coefficient) constraints via a suppression factor, $\mathcal{S}$, which is determined as a ratio of the extracted limit at a given $c_n/c_p$ to the limit at the
extracted 90\% C.L.\ limit evaluated in the isospin-conserving case: 
\begin{equation}
\mathcal{S}(m, c_n/c_p) \equiv \frac{\sigma^{90\% {\rm C.L.}}_{\chi N}(m, c_n/c_p)}{\sigma^{90\% {\rm C.L.}}_{\chi N}(m)\vert_{c_n=c_p}}.
\label{eq:suppresion_def}
\end{equation}
This suppression factor is determined by explicitly re-deriving the experimental
constraints with a new dark matter signal model, accounting for the energy recoil thresholds, efficiencies, and background contributions of the experiments under consideration as described in Sec.~\ref{sec:exp_setup}.

In Figure~\ref{fig:1d-suppressions_SI}, the suppression factor for the DarkSide-20k-like argon limit and the LZ-like xenon experimental limit is plotted for neutron-to-proton coupling ratios between $-1$ and $+1$ for momentum-independent operators $\mathcal{O}_{1}$ and $\mathcal{O}_{8}$ and for the momentum-dependent operator $\mathcal{O}_{11}$. We present the inverse of the $\mathcal{S}$ suppression factor, meaning lower values correspond to a stronger suppression of the limits.
The suppression behaviour is strongly dependent on the value of the neutron-to-proton coupling ratio, with target and operator-dependent values where experimental sensitivity
to dark matter would be maximally suppressed.
For the momentum-independent $\mathcal{O}_{1}$ operator in a xenon target, this is around $c_{n}/c_{p}=-0.7$, compared to an argon target where this peak occurs around $c_{n}/c_{p}=-0.8$.
The dependence of the suppression factor on the coupling ratio near the region of peak suppression is also affected by the form factors used in the calculation, as could be inferred by Figure~\ref{fig:FormFactor_difference}, we explicitly show the effects in Figure~\ref{fig:1dSuppressionOLDvsNEWFF} in the Appendix for the $\mathcal{O}_{1}$ operator.

\begin{figure*}[htb]
    \centering
    \includegraphics[width=0.8\textwidth]{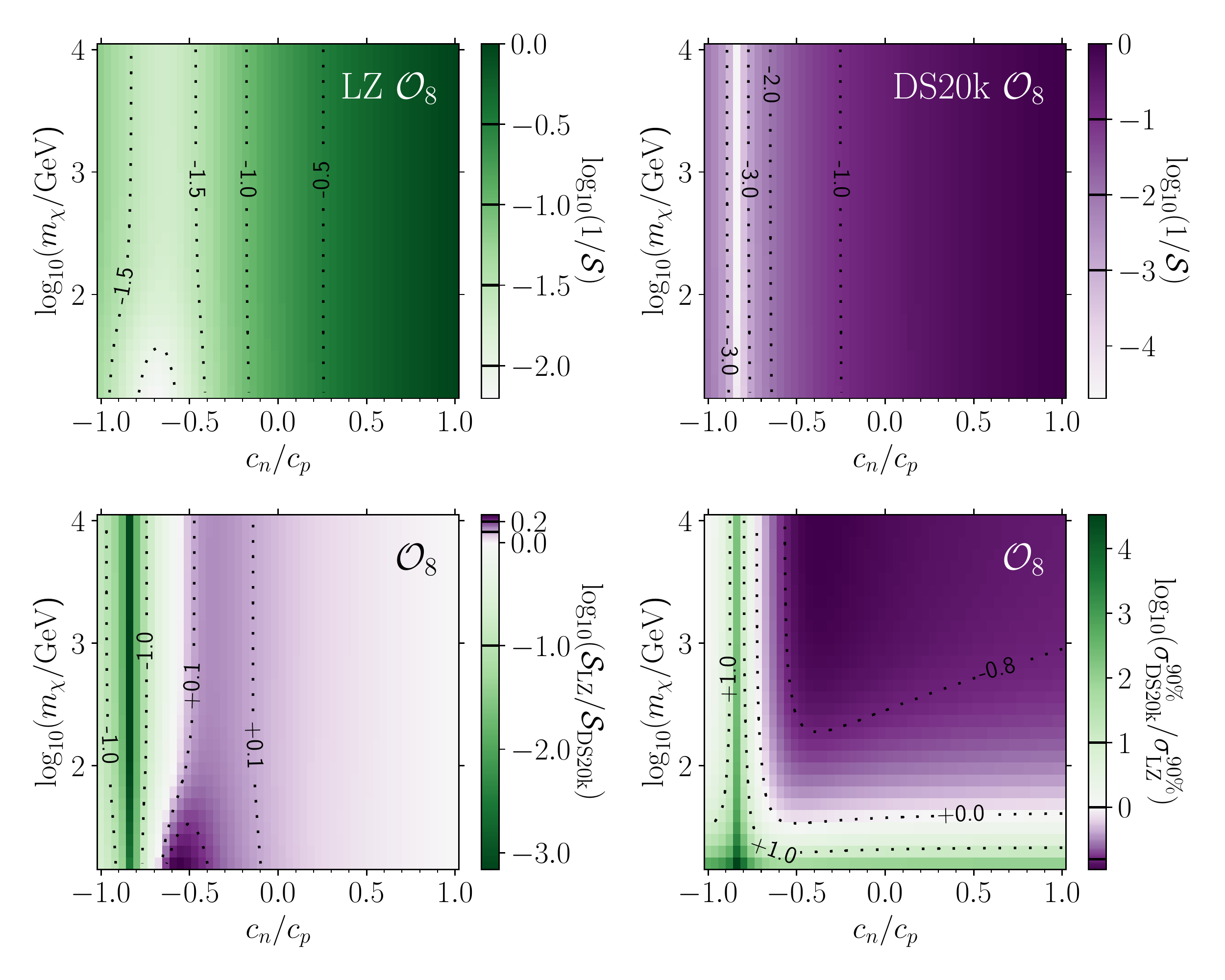}
    \caption{\textbf{Top row:} Inverse of the experimental suppression factor for an LZ-like experiment (top left) and a DarkSide-20k-like experiment (top right) as a function of neutron-to-proton coupling ratio and dark matter candidate mass, for the $\mathcal{O}_8$ operator. Large negative values indicate a strong suppression of the experimental limit.
    \textbf{Bottom left:} Ratio of suppression in LZ-like to DarkSide-20k-like experiment. Negative $z$-values (green regions) identify regions where LZ would suffer from less suppression than Darkside-20k, with positive values (purple regions) identifying the opposite.
    \textbf{Bottom right:} : Ratio of the projected 90\% exclusion limit in DarkSide-20k to LZ, where here negative values (green regions) indicate where the full exposure of an LZ-like detector would be the more sensitive experiment for a particular mass and coupling ratio while positive values (purple) indicates where DarkSide-20k would be the most sensitive.
    \label{fig:2dSuppression}
    }
\end{figure*}

For $\mathcal{O}_{1}$, the dark matter mass dependence of the suppression is minimal. Still, results for the $\mathcal{O}_{8}$ and $\mathcal{O}_{11}$ operators illustrate significant mass dependence of the coupling ratio value corresponding to maximum suppression as well as the amount of suppression. This is particularly true for xenon and much less so for argon.
Given this mass-dependent effect, we also present in the top row of Figure~\ref{fig:2dSuppression} a two-dimensional version of these plots (here only $\mathcal{O}_8$, for illustration),
again redetermining the projected experimental constraints for each experiment at each mass and coupling ratio point,
to illustrate how the limit suppression relative to the isospin-symmetric scenario
varies as a function of dark matter candidate mass and coupling ratio. Again we present the inverse of the suppression factor in the top row $1/\mathcal{S}$, where large negative values (lighter colours) represent a stronger suppression.
We also display the ratio of the suppression in argon (for a DarkSide-20k-like detector) and xenon (for an LZ-like detector), shown in the bottom left pane of Figure~\ref{fig:2dSuppression}. 
This metric provides information about how the experimental limits for different targets will be affected in comparison to each other. For these plots, green regions denote points where the xenon LZ-like limit is suppressed more strongly, and in the purple areas, the argon DarkSide-20k-like experimental limit is suppressed to a greater degree.
Finally, we also present in the bottom right panel the ratio of the exclusion limits, to demonstrate the mass and coupling ratio combinations for which DarkSide-20k or LZ, with their full exposures, would be projected to be the most sensitive. Rather than being a measure of the relative suppressions in Ar compared to Xe regardless of the nominal sensitivity, this gives a sense of which target is more sensitive for the specific exposures and setup of DarkSide-20k and LZ.

\begin{figure*}[htbp]
    \centering
    \includegraphics[width=\textwidth]{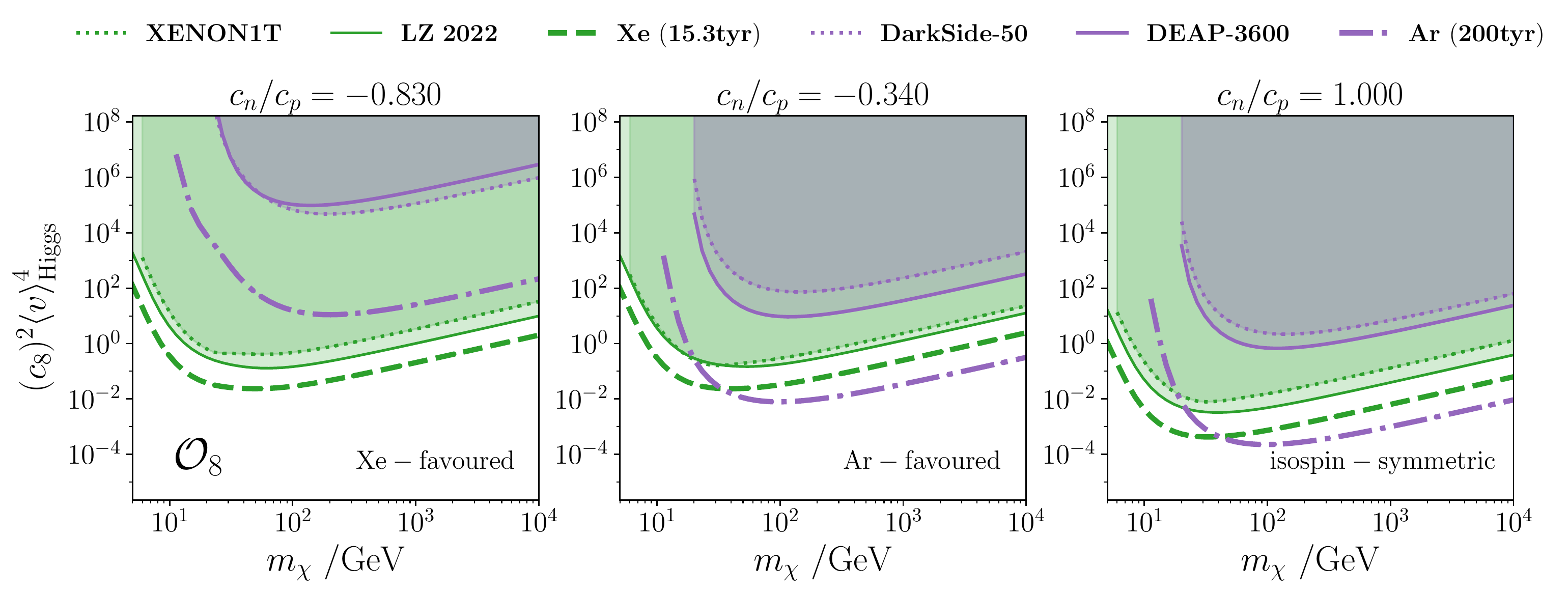}
    \caption{Current LXe and LAr dark matter exclusion regions [shaded] and future projections for two near-future liquid noble direct detection experiment sensitivities (a 200 tonne-year argon experiment and a 15 tonne-year xenon experiment) [dashed] 
    as a function of nucleon -- dark matter EFT interaction coefficient $c_8$
    for the $\mathcal{O}_8$ operator a xenon-favoured model scenario (left), an argon-favoured model scenario (middle), and for the standard isospin-conserving case (right).
    }
    \label{fig:sens-compareratios-O8}
\end{figure*}

We highlight that incorporating the efficiency (and the model used) affects the signal spectra and projected experimental limits, therefore the maximal suppression rate and $c_n/c_p$ value are also modified. 
We have explored the effect of efficiencies by switching on and off certain experimental inputs and find that the effect on the suppression is negligible and thus the results presented here are insensitive to changes in the expected performance of these detectors.

The equivalent to Figure~\ref{fig:2dSuppression} for four other operators ($\mathcal{O}_1$, $\mathcal{O}_3$, $\mathcal{O}_5$, and $\mathcal{O}_{11}$) can be found in the Appendix (Figure~\ref{fig:app_2d_supp_op}).
We note that the result for $\mathcal{O}_1$ is fairly mass-independent in terms of the $c_n/c_p$ value corresponding to maximum suppression, although the maximum amount of suppression reduces with increasing dark matter mass. The equivalent of the bottom right plot of Figure~\ref{fig:2dSuppression} for the other four operators, where we present the ratio of the exclusion limits, is presented in Figure~\ref{fig:app_2d_xsecratio} of the Appendix.
For completeness, two-dimensional suppression factors for operators for which xenon is sensitive, but argon is not due to its nuclear spin being zero ($\mathcal{O}_4$, $\mathcal{O}_6$, $\mathcal{O}_7$, $\mathcal{O}_9$, and $\mathcal{O}_{10}$) are also presented in the Appendix (Figure~\ref{fig:app_2d_supp_xe}) and are shown to be relatively mass independent with maximum suppression at $c_n/c_p$ close to zero. For $\mathcal{O}_4$ the maximum suppression occurs at $c_n/c_p=-0.03$. 

Additionally, we evaluate the experimental limits for the operators separately for the current and future detectors as we did for the models in the previous section. In Figure~\ref{fig:sens-compareratios-O8} we present the experimental limits for operator $\mathcal{O}_8$ evaluated as an exclusion limit on the operator coefficient $c_8$. The xenon-favoured and argon-favoured $c_n/c_p$ values were chosen as the coupling ratio that maximised/minimised the suppression ratio shown in Figure~\ref{fig:2dSuppression} at a benchmark mass of 100~GeV, roughly the mass at which the experiments are most sensitive. The equivalent to Figure~\ref{fig:sens-compareratios-O8} for other operators can be found in the Appendix (Figures~\ref{fig:sens-compareratios-O1O3} and \ref{fig:sens-compareratios-O5O11}).

Our aim here is that the results presented will be useful for those who want to estimate the effects of specific isospin-violating models. For example, say one had a model that predicted a $\mathcal{O}_8$ nuclear response with a neutron-to-proton coupling factor of $c_n/c_p=-0.5$, one could take our projected sensitivities for the isospin-symmetric case on the right-most panel of Figure~\ref{fig:sens-compareratios-O8}, and use the top two panels of Figure~\ref{fig:2dSuppression} to determine the value of $\mathcal{S}$. 
For $c_n/c_p=-0.5$, and the $\mathcal{O}_8$ interaction as an example, a $100\,{\rm GeV}$ dark matter candidate has suppression factors of $\mathcal{S}\approx 35$ and $\approx 30$ for xenon and argon respectively. Using the respective limits and the value for $\mathcal{S}$ as in Eq.\,\eqref{eq:suppresion_def}, one can determine how future LZ and DarkSide-20k experiments will constrain the model in question. The isospin-symmetric limits at $100\,{\rm GeV}$ are
$c^2_{8}\langle v \rangle^4_\mathrm{Higgs}=8.0\times10^{-4}$ and $2.0\times10^{-4}$ for LZ and DarkSide-20k in the isospin symmetric case, so for the spin-dependent $\mathcal{O}_8$ operator and $c_n/c_p=-0.5$ case, the limits 
would be $2.8\times 10^{-2} $ and $6.0\times 10^{-3}$ for LZ and DarkSide-20k, respectively. 
Given the insensitivity of our calculated suppression ratios to experimental features such 
as efficiency effects and exposure, these values could also be used to estimate 
the sensitivity of future xenon and argon experiments to isospin-violating dark matter.

\section{Conclusion}
\label{sec:conclusion}

In this article, we have revisited the case of isospin-violation in dark matter interactions with the SM. This occurs in a substantial number of models and has the potential to significantly suppress the recoil rate in direct detection and thus weaken the apparent current and future projected constraints by many orders of magnitude. We pay particular attention to the crucial complementarity of xenon and argon-target noble liquid detectors which, in the coming decade, will be the most sensitive for $m_\chi\geq 10$~GeV for S1+S2 analyses and $m_\chi\geq 50$~MeV for S2-only analyses. In this work, we have focused on the S1+S2 analysis strategy, but note that many of the qualitative results would extend down to 50~MeV when an ionisation analysis is incorporated. 
Additionally we dedicated some effort to highlight the importance of performing a spectral analysis for isospin-violating models. We showed that the magnitude of suppression can be altered by the dark matter mass, experimental configuration and nuclear form factors. We have tried to account for variations in the former two throughout this work, and have simply exhibited the importance of the latter, hoping to motivate more dedicated work. 

To explore these points in a concrete way we have primarily focused on two gauge-invariant effective models of dark matter, which are inspired by Dirac and Majorana dark matter candidates that couple to the SM via vector currents. These two examples we refer to as SI and SD models, due to their dominant operator responses. We have performed new analyses of the best available public data (from the XENON1T, LZ, DarkSide-50, and DEAP-3600 experiments) to set new constraints on isospin-violating dark matter and determined the effects in the case where one target element is particularly favoured. We do the same for future sensitivity projections in LZ-like and DarkSide-20k-like experimental setups. Interestingly we find that, due to gauge invariance, the SD model, which is canonically considered to be invisible to argon target experiments, is not. In fact, even in the isospin-conserving case, our projections show DarkSide-20k will be more sensitive than xenon for SD interactions for $m_\chi > 100$~GeV. 

Furthermore, we have performed the same analysis on the more model-independent and comprehensive NREFT operators developed for direct detection, in the hope that those in the model-building community can apply our projections to their isospin-violating models. In Section~\ref{sec:results} we only presented one operator fully, $\mathcal{O}_8$, in order to describe our presentation and how one may use our results. We have provided our results for operators $\mathcal{O}_{1-11}$ in the Appendix.  

\section*{Data availability statement}
Data available from the main results of this paper is available at Ref.~\cite{Price2023}.

\acknowledgements
AC is supported by the grant ``AstroCeNT: Particle Astrophysics Science and Technology Centre" carried out within the International Research Agendas programme of the Foundation for Polish Science financed by the European Union under the European Regional Development Fund.
DP is supported by the UKRI's Science and Technology Facilities Council (STFC) under grants ST/M005437/1 and ST/N000374/1 and by the University of Manchester. 
ES is supported by STFC and a George Rigg Presidential PhD scholarship from the University of Manchester.
We thank Dorian Amaral, David Cerde\~no, Aaron Manalaysay, and Felix Kahlhoefer for useful discussions. 

\FloatBarrier

\twocolumngrid
\bibliography{isospinbib}

\newpage
\onecolumngrid
\appendix
\section*{Appendix}

\subsection{Limit suppression for the $\mathcal{O}_1$ interaction with neutron-to-proton coupling ratio and form factor variations} 
\label{app:formfact_suppression}

\begin{figure*}[htbp]
    \centering
    \includegraphics[width=0.6\textwidth]{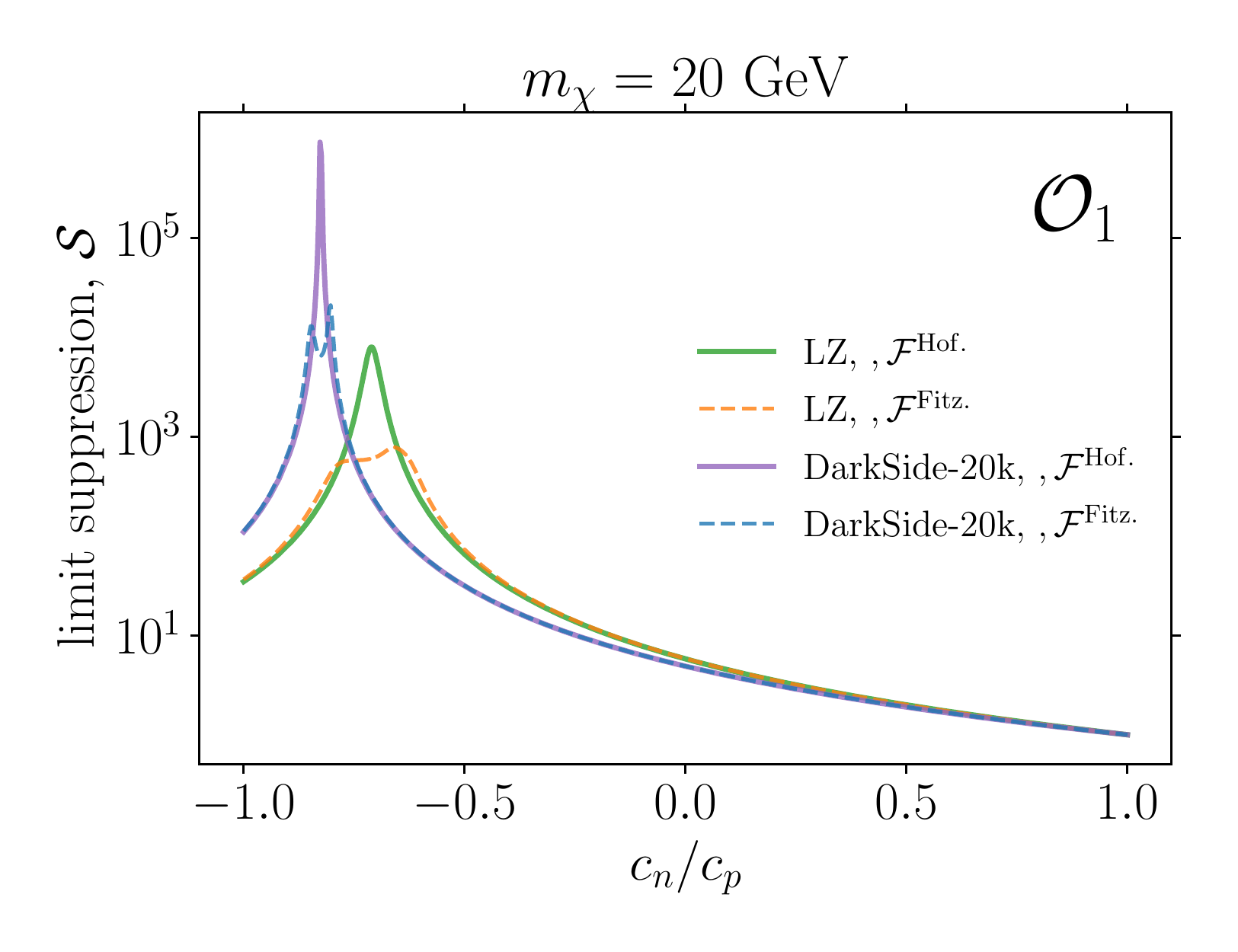}
    \caption{Suppression of the experimental constraints on the dark matter -- nucleon interaction strength with neutron-to-proton coupling ratio compared to the constraints for the isospin-conserving scenario at a representative dark matter mass of 20~GeV,
    for the $\mathcal{O}_{1}$ interaction for the two form factors considered in this paper. Efficiency effects and energy window constraints for the LZ and DarkSide-20k detectors have been incorporated. 
    }
    \label{fig:1dSuppressionOLDvsNEWFF}
\end{figure*}

Figure~\ref{fig:1dSuppressionOLDvsNEWFF} illustrates how the experimental constraints on dark matter -- nucleon interaction strength are suppressed as a function of the coupling ratio $c_n/c_p$ to the $\mathcal{O}_1$ interaction for different targets at a dark matter candidate mass of 20~GeV, and how these are substantially affected by the choice of form factors $\mathcal{F}^{\rm Hof.}$ and $\mathcal{F}^{\rm Fitz.}$.

\subsection{Mass and coupling dependence of suppression of experimental constraints}
\label{app:2D}
\begin{figure*}[htbp]
    \centering
    \includegraphics[width=\textwidth]{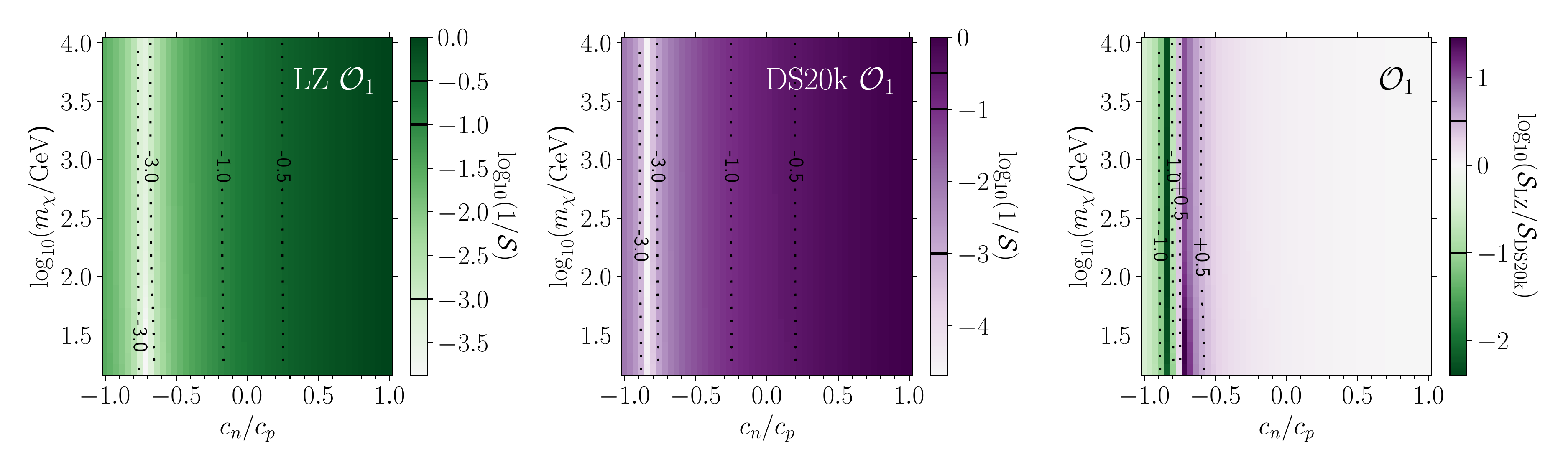}
    \includegraphics[width=\textwidth]{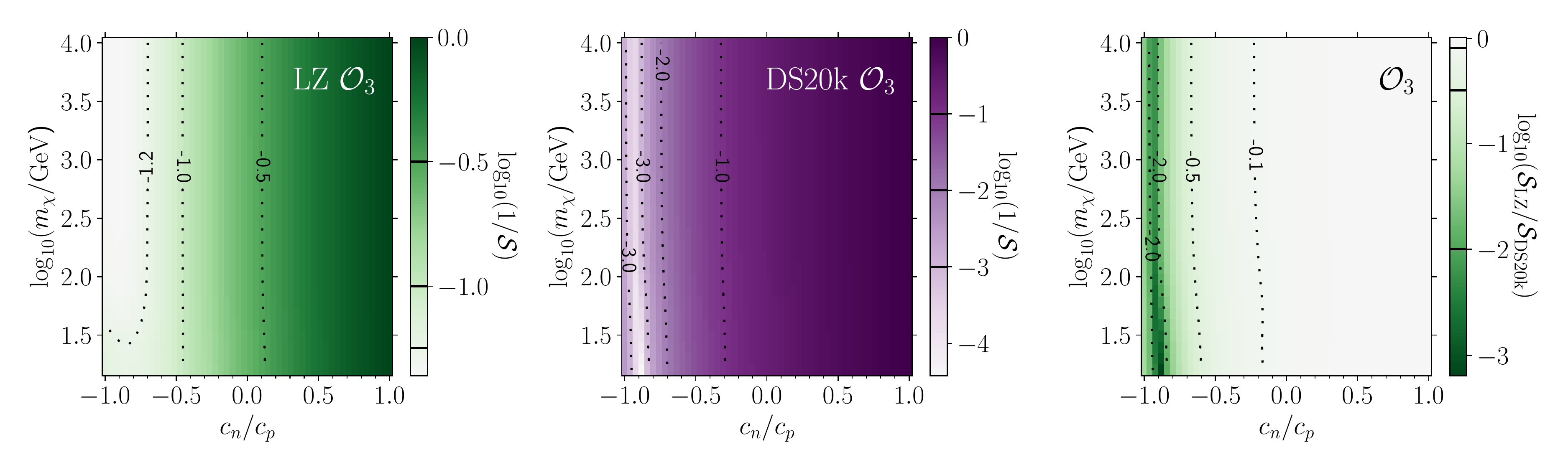}
    \includegraphics[width=\textwidth]{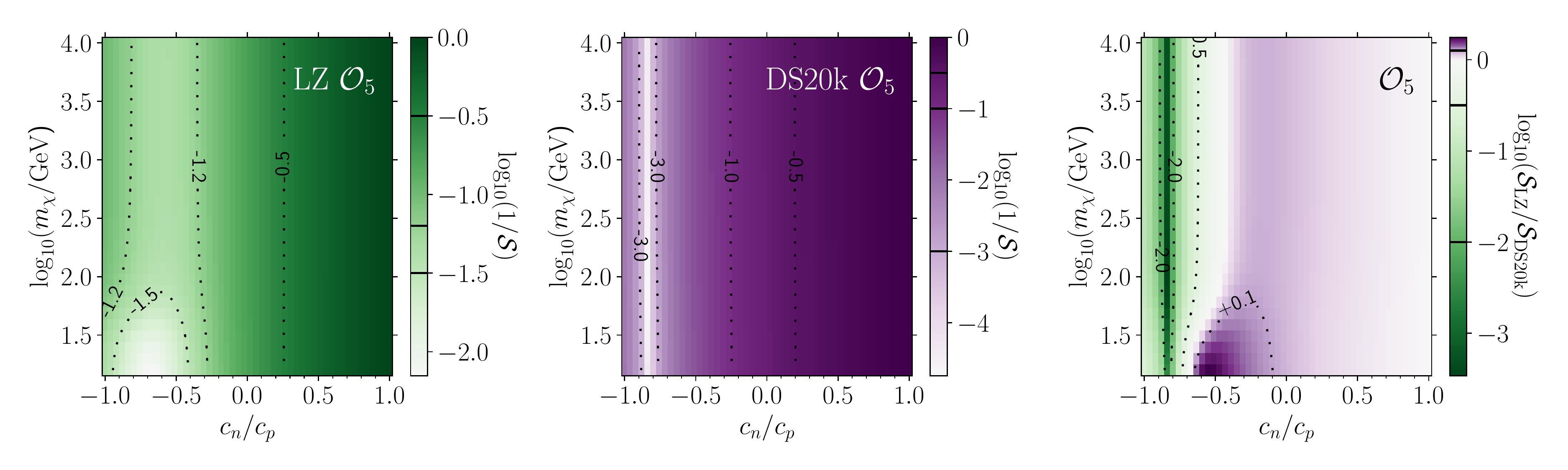}
    \includegraphics[width=\textwidth]{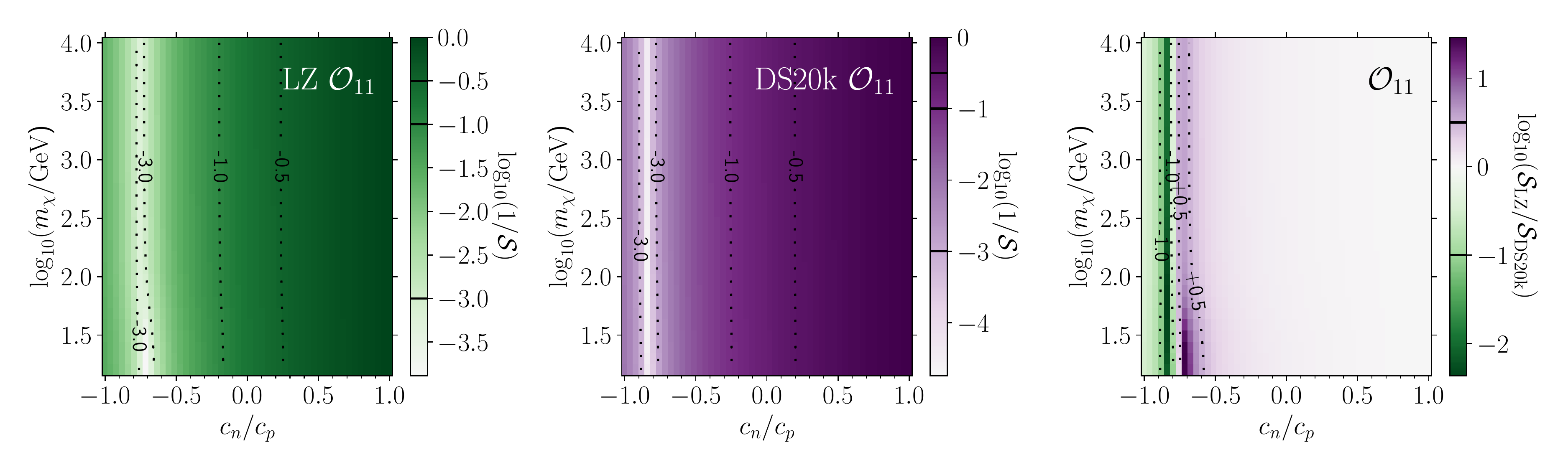}
    \caption{Inverse suppression factor of experimental constraints on dark matter -- nucleon interaction strength as a function of neutron-to-proton coupling ratio and dark matter mass for an LZ-like experiment (left), DarkSide-20k-like experiment (middle) and the ratio of DarkSide-20k-like to LZ-like (right) for the  $\mathcal{O}_1$, $\mathcal{O}_3$, $\mathcal{O}_5$, and $\mathcal{O}_{11}$ operators.
    }
    \label{fig:app_2d_supp_op}
\end{figure*}

\begin{figure*}[htbp]
    \centering
    \includegraphics[width=\textwidth]{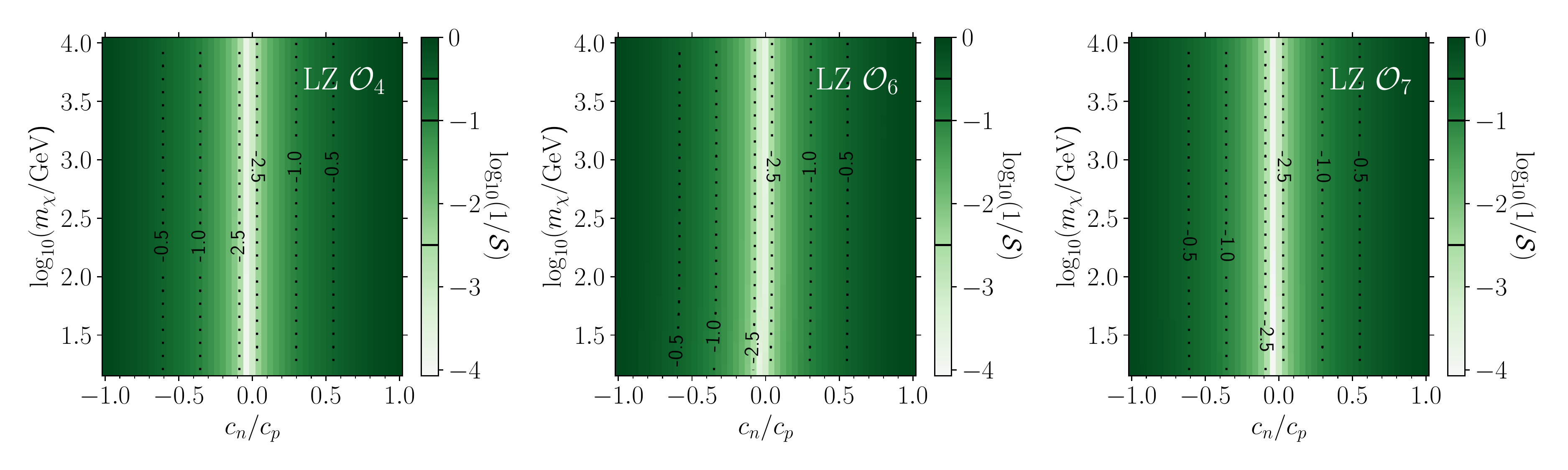}
    \includegraphics[width=0.66\textwidth]{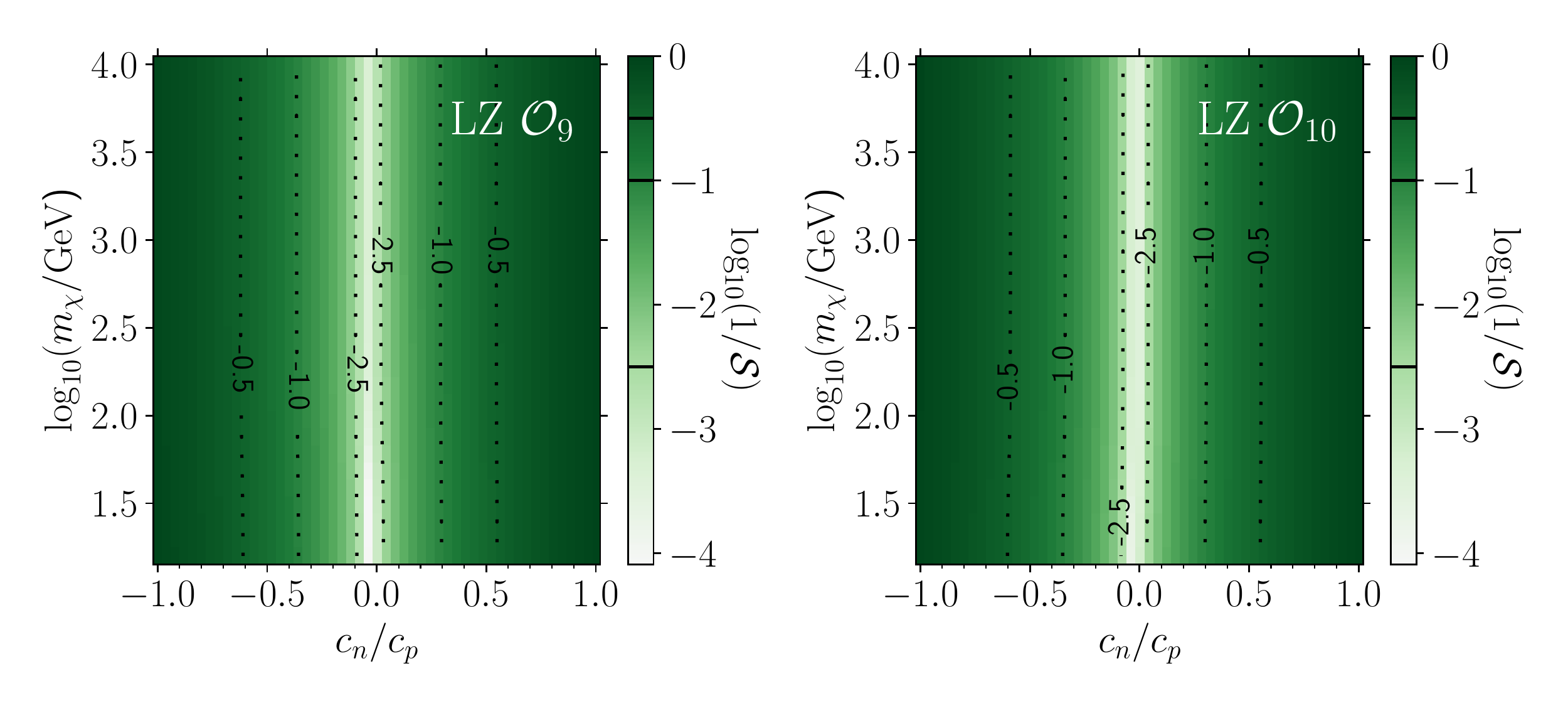}
    \caption{Experimental constraints on dark matter -- nucleon interaction strength as a function of neutron-to-proton coupling ratio and dark matter mass for an LZ-like experiment, for the $\mathcal{O}_4$, $\mathcal{O}_6$, $\mathcal{O}_7$, $\mathcal{O}_{9}$, and $\mathcal{O}_{10}$ operators.
    }
    \label{fig:app_2d_supp_xe}
\end{figure*}

\begin{figure*}[htbp]
    \centering
    \includegraphics[width=0.8\textwidth]{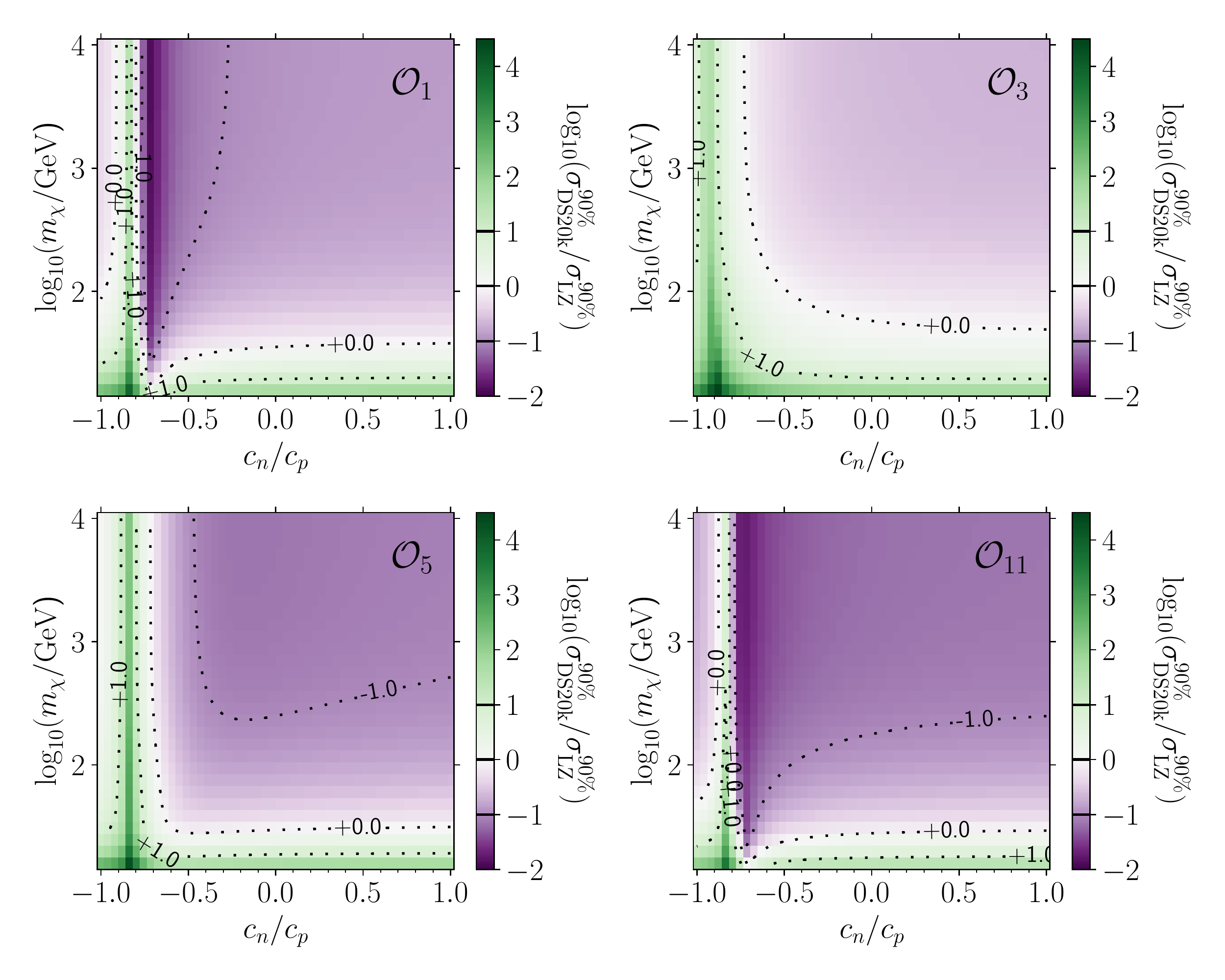}
    \caption{Ratio of the projected experimental constraints on dark matter -- nucleon interaction strength as a function of neutron-to-proton coupling ratio and dark matter mass for an LZ-like experiment versus a DarkSide-20k-like experiment with their full projected exposures for the  $\mathcal{O}_1$, $\mathcal{O}_3$, $\mathcal{O}_5$, and $\mathcal{O}_{11}$ operators.
    }
    \label{fig:app_2d_xsecratio}
\end{figure*}

Figure~\ref{fig:app_2d_supp_op} and~\ref{fig:app_2d_supp_xe} show two dimensional distributions illustrating how experimental constraints on dark matter -- nucleon interaction strength are affected by both $m_{\chi}$ and $c_n/c_p$. We show this dependence for multiple NREFT operators that are not included in the main body of the text (Figure~\ref{fig:app_2d_supp_op}) and for the operators only present
for xenon targets and not for argon (Figure~\ref{fig:app_2d_supp_xe}).
Additionally, we show the ratio of the projected exclusion limit in DarkSide-20k and LZ in the same parameter space, for four operators in Figure~\ref{fig:app_2d_xsecratio}.

\FloatBarrier
\subsection{Current and projected dark matter -- nucleon interaction experimental sensitivities}
\label{app:results}
\begin{figure*}[htbp]
    \centering
    \includegraphics[width=\textwidth]{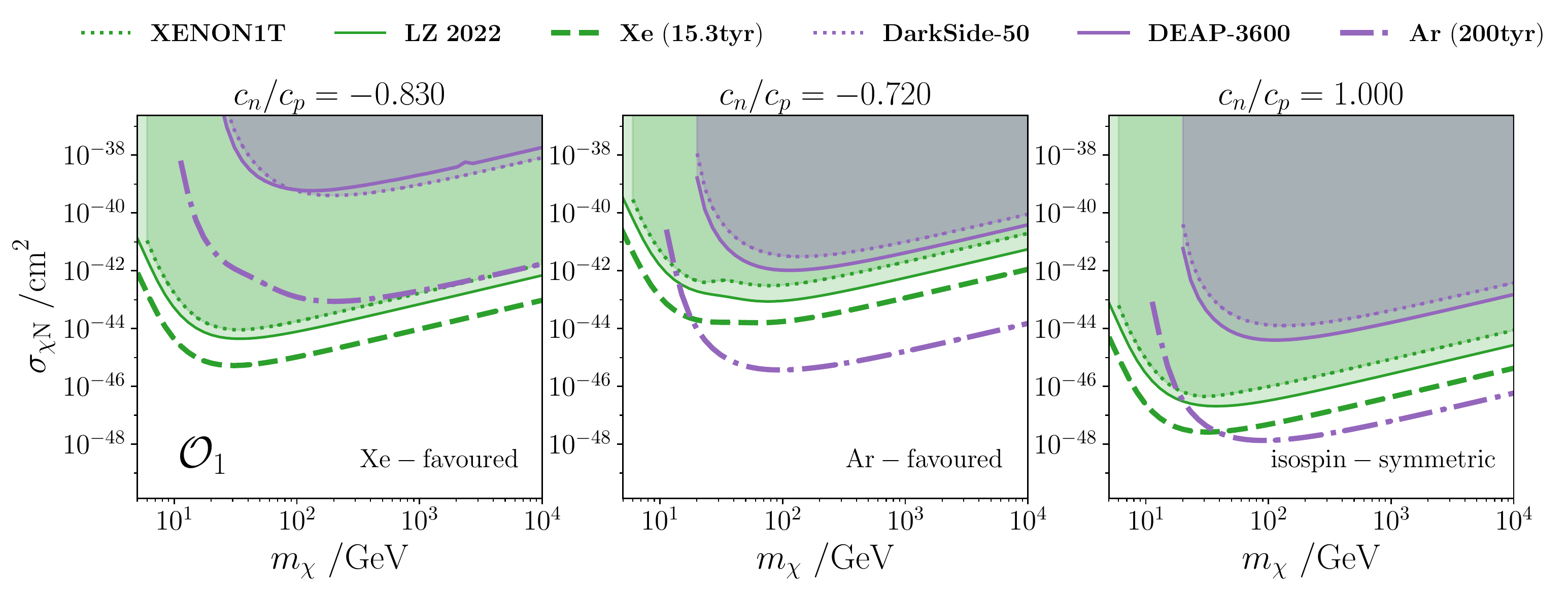}
    \includegraphics[width=\textwidth]{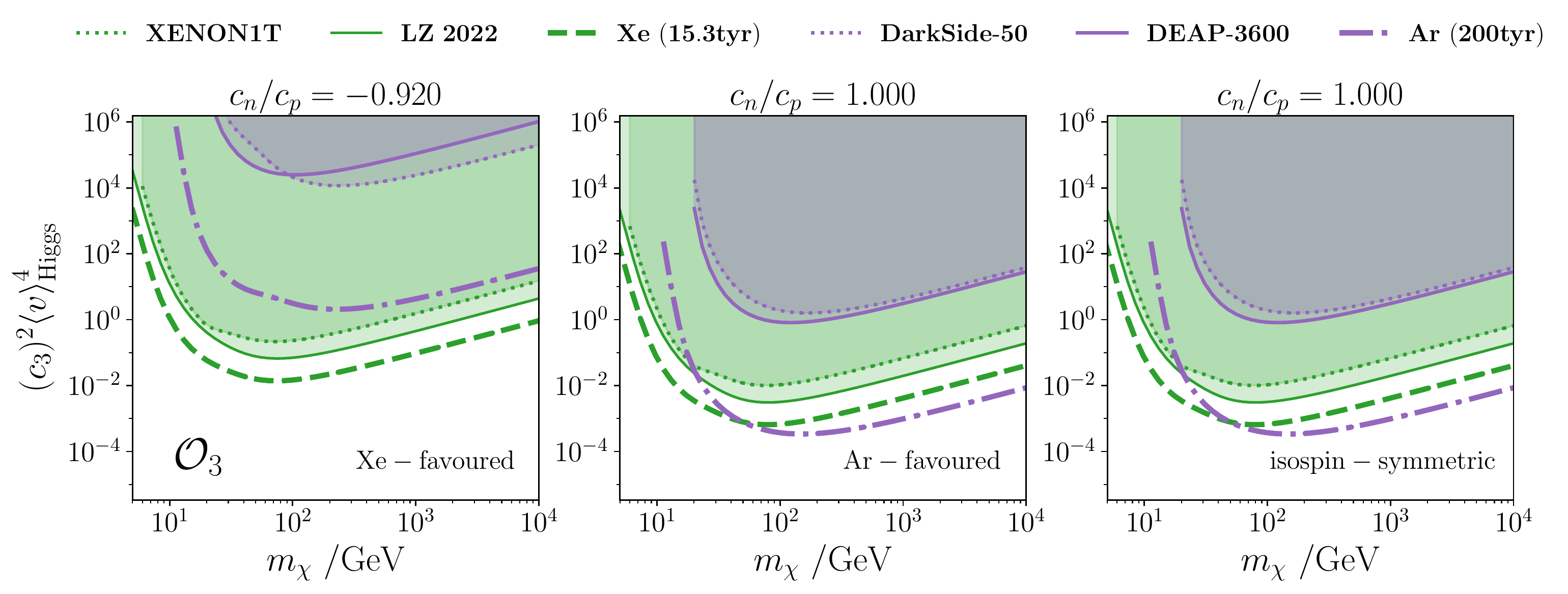}
    \caption{Current LXe and LAr dark matter exclusion regions [shaded] and future projections for two near-future liquid noble direct detection experiment sensitivities (a 200 tonne-year argon experiment and a 15 tonne-year xenon experiment) [dashed] 
    as a function of spin-independent dark matter -- nucleon interaction cross-section
    for the \textbf{(top)} $\mathcal{O}_1$ and as  a function of the $c_3$ coefficient \textbf{(bottom)} for the $\mathcal{O}_3$ operator in a xenon-favoured model scenario (left), an argon-favoured model scenario (middle), and for the standard isospin-conserving case (right).
    }
    \label{fig:sens-compareratios-O1O3}
\end{figure*}

\begin{figure*}[htbp]
    \centering
    \includegraphics[width=\textwidth]{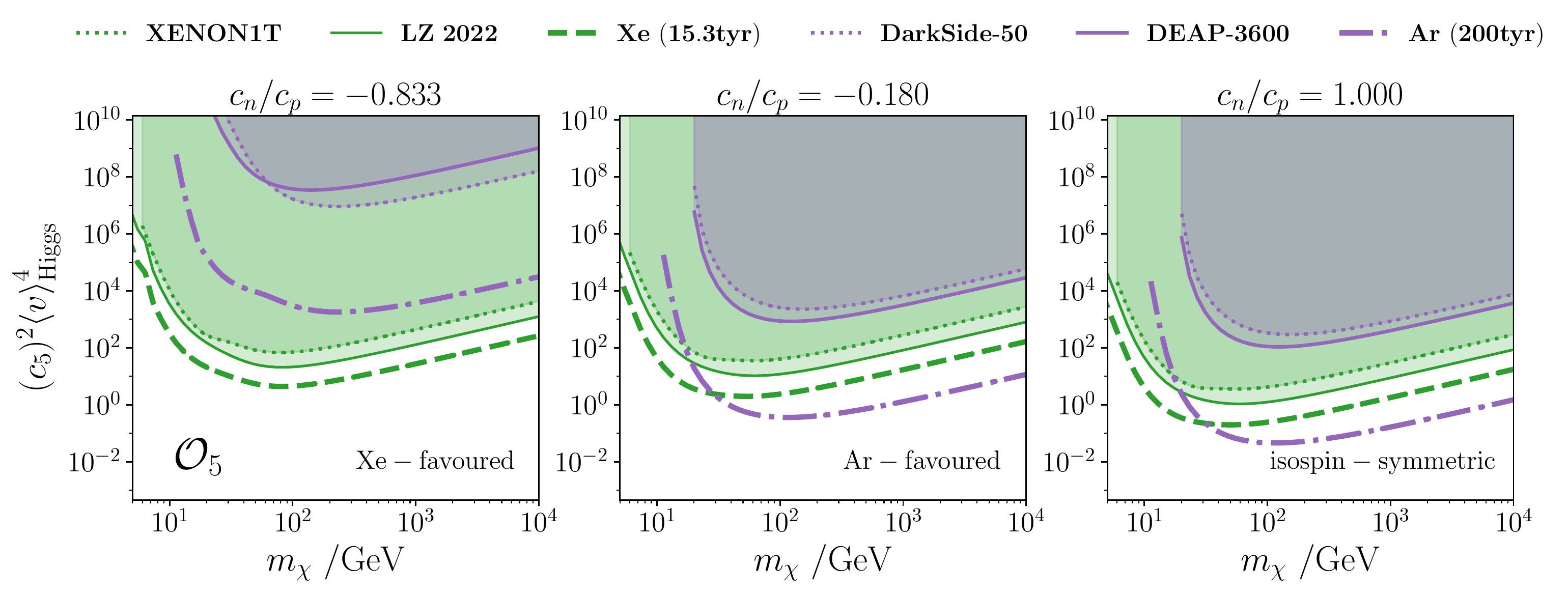}
    \includegraphics[width=\textwidth]{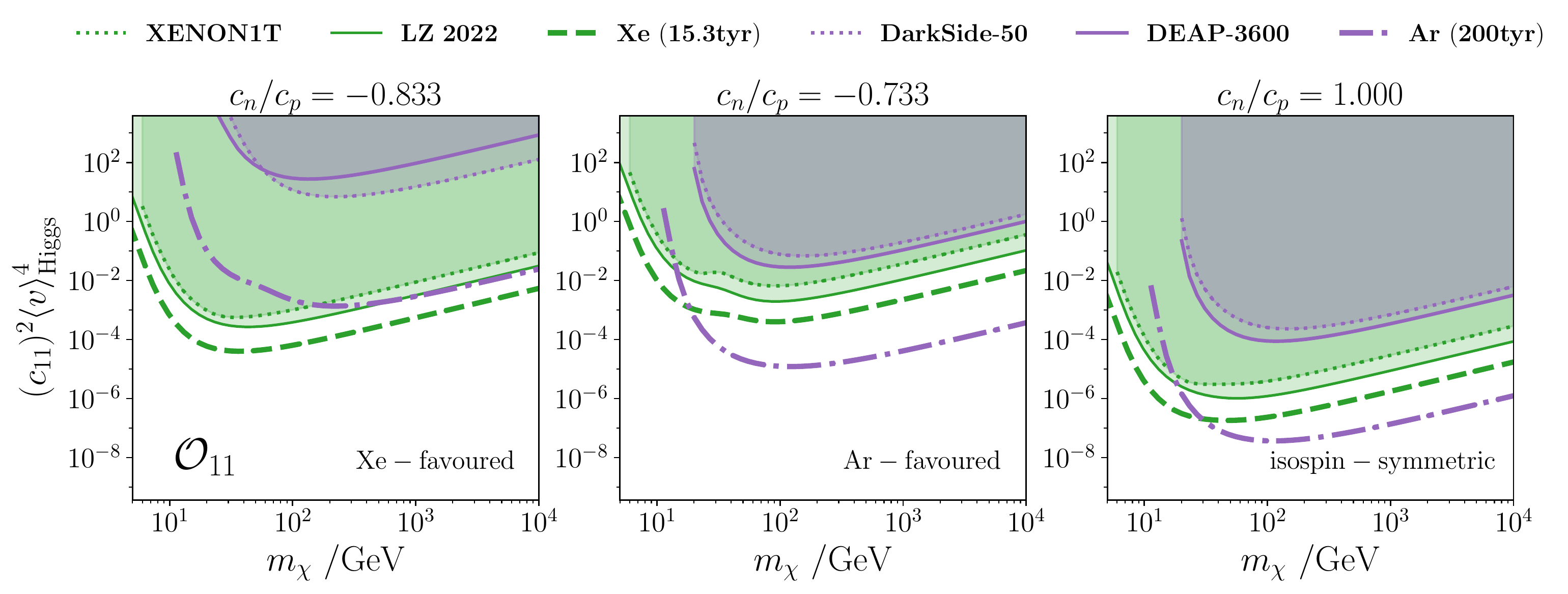}
    \caption{Current LXe and LAr dark matter exclusion regions [shaded] and future projections for two near-future liquid noble direct detection experiment sensitivities (a 200 tonne-year argon experiment and a 15 tonne-year xenon experiment) [dashed] 
    as a function of nucleon -- dark matter EFT interaction coefficients $c_5$ \textbf{(top)} and $c_{11}$ \textbf{(bottom)} for the  $\mathcal{O}_5$ and $\mathcal{O}_{11}$ operators in a xenon-favoured model scenario (left), an argon-favoured model scenario (middle), and for the standard isospin-conserving case (right).
    }
    \label{fig:sens-compareratios-O5O11}
\end{figure*}

Figures~\ref{fig:sens-compareratios-O1O3} and~\ref{fig:sens-compareratios-O5O11} illustrate the current and projected xenon and argon limits for various nucleon--dark matter EFT interactions as a function of interaction rate / interaction coefficient and dark matter candidate mass, 
modified assuming xenon favoured (left panel), argon favoured (middle) and isospin-conserving (right panel) configurations. 

For the $\mathcal{O}_1$ operator, the coefficient value can be related to the cross-section at zero momentum transfer, using the following relation:
\begin{equation}
\sigma_{\chi N}=\frac{\mu_{\chi N}^{2}}{\pi }{c}_{1}^{2}.
\end{equation}
For all other operators, we present the results in terms of the NREFT coefficients directly, 
as for momentum-dependent operators in particular the zero-momentum transfer cross-section relationship above does not correspond to a physically meaningful quantity. We present results in terms of the coupling coefficient to the proton squared, which in the case of a single operator and isospin conservation is directly related to the rate of dark matter events.

\end{document}